# From Stacking Disorder to Cubic Order: Ice Crystallization from Deeply Supercooled Water

Yulin Lin[1,2], Weimin Guo[3], Suvo Banik[1], Tao Zhou[1]*, Thomas E. Gage[1], Lei Yu[1], Maksim A. Sultanov[1], Martin Holt[1], Subramanian Sankaranarayanan[1]*, Peng Zhang[3]*, Ilke Arslan[1], Jianguo Wen[1*]

**Affiliations:**
[1]Center for Nanoscale Materials, Argonne National Laboratory, Lemont, Illinois, United States

[2]Institute for Advanced Studies, College of Chemistry and Molecular Sciences, Wuhan University, Wuhan, P. R. China

[3]Institute of Refrigeration and Cryogenics, Shanghai Jiao Tong University, Shanghai, P.R. China

***Corresponding authors**: Tao Zhou (tzhou@anl.gov); Subramanian Sankaranarayanan (ssankaranarayanan@anl.gov); Peng Zhang (zhangp@sjtu.edu.cn); Jianguo Wen (jwen@anl.gov)

## Abstract:

Crystallization far from equilibrium can generate morphologies that defy classical crystal habits, yet the microscopic mechanisms linking atomic-scale disorder to emergent macroscopic order remain elusive. Here we use in situ cryogenic transmission electron microscopy with a membrane-encapsulated microdroplet platform to directly visualize the freezing of deeply supercooled water at molecular resolution. We show that homogeneous nucleation produces stacking-disordered ice composed of mixed hexagonal and cubic sequences, in which cubic ice initially exists only as isolated monolayers. The gradual thickening of these cubic layers constitutes the key kinetic mechanism that governs the entire crystallization pathway. As thickening proceeds, nanoscale, defect-free cubic ice germs nucleate on the basal planes of the disordered lattice. These faceted cubic germs act as facet-registered kinetic seeds that enforce cubic twinning and sequentially multiply growth branches. This kinetic pathway reproducibly generates robust eight-branched dendrites with global cubic (octahedral) symmetry, even though each branch remains highly stacking-disordered. At later stages, latent heat release drives a crossover to the thermodynamically favored hexagonal phase; remarkably, the pre-established global cubic symmetry is retained. These results reveal how strong kinetic driving forces convert microscopic disorder into emergent macroscopic symmetry, providing a general framework for understanding and controlling rapid crystallization far from equilibrium.

**Key words**: Ice crystallization, Stacking disorder, Kinetic growth/Rapid growth, Cubic–hexagonal polytypism, In situ cryogenic TEM

## Main Text

Crystallization far away from equilibrium plays a central role in materials synthesis [1–3], climate science [4], pharmaceuticals [5], to food processing [6]. Under extreme kinetic driving forces, crystal growth can deviate markedly from thermodynamic equilibrium, producing structures and morphologies that cannot be inferred from equilibrium phase diagrams alone [7]. Despite its broad importance, the microscopic pathways by which atomic-scale disorder generated during nucleation evolves into emergent macroscopic crystal order remain poorly understood. This gap largely arises because rapid solidification processes are difficult to probe directly at the relevant spatial and temporal scales.

Ice crystallization from deeply supercooled water provides a paradigmatic system for investigating nonequilibrium crystallization. Large undercooling leads to explosive nucleation and ultrafast growth, frequently accompanied by extensive stacking disorder and complex twinning [8–15]. X-ray diffraction studies have established that ice nucleated from supercooled water commonly adopts a stacking-disordered structure composed of mixed hexagonal ($I_h$) and cubic ($I_c$) layers [16–21]. However, diffraction-based approaches inherently average over large volumes and therefore lack the spatial resolution required to resolve the nucleation core or the earliest growth stages. Molecular simulations have provided valuable atomistic insight into ice nucleation [22–29], but are typically limited to very early events and cannot capture the multiscale evolution of crystal morphology during rapid growth.

A striking and long-standing puzzle is that ice nucleated from deeply supercooled water, despite being strongly stacking-disordered at the atomic scale, often develops macroscopic growth morphologies characterized by well-defined twinning angles of 70.5° (or the complementary 109.5°), which are hallmarks of cubic symmetry [10,11,13,14,30]. This raises a fundamental question: how can a lattice that is locally disordered organize into a globally ordered, symmetry-governed structure? More broadly, the challenge of controlling emergent microstructure from highly disordered building blocks is not unique to ice. Stacking disorder and polytypism are widespread in rapidly synthesized materials, including SiC, diamond, GaN, Mg-based alloys, and transition metals, where nonequilibrium growth often stabilizes mixed cubic–hexagonal sequences [31–33]. In these systems, disorder can act as a functional microstructural motif, strongly influencing mechanical, electronic, thermal, and magnetic properties. Yet, despite extensive empirical evidence linking stacking disorder to enhanced performance, a unifying understanding of how local disorder organizes across length scales—and how it can be harnessed to produce persistent, symmetry-governed architectures—remains elusive.

Addressing this problem requires direct, molecular-scale observation of homogeneous ice nucleation and growth under deep supercooling—a long-standing experimental challenge due to the incompatibility of liquid water with the high vacuum of electron microscopy. While atomic- or molecular-resolution imaging has been achieved in membrane liquid cells at ambient or cryogenic conditions, true molecular-resolution imaging of dynamic nucleation and growth processes of ice in cryogenic membrane liquid cells has not yet been demonstrated. While atomic-resolution imaging has been achieved in membrane liquid cells at room temperature [34], and molecular-resolution imaging has been realized for vapor-grown ice in microscope vacuum [35] or in cryogenic liquid cell [36], true molecular-resolution imaging of the homogeneous nucleation and growth of ice directly from supercooled liquid water has not yet been demonstrated.

Here we overcome this limitation by combining cryogenic transmission electron microscopy with a membrane-encapsulated microdroplet platform that enables repeatable, in situ imaging of ice formation at −40 °. This approach allows precise localization of nucleation centers and tracking of structural evolution across multiple length scales during rapid freezing. We show that ice initially nucleates as stacking-disordered ice with low cubicity, but that nanoscale cubic ice germs emerge during growth and act as facet-registered, kinetic symmetry-enforcing seeds. These germs drive successive twinning events across equivalent {111} orientations, producing a robust octahedral dendritic morphology with global cubic symmetry, even though each branch remains stacking-disordered. Our results reveal a previously unrecognized nonequilibrium crystallization pathway in which facet-governed kinetic constraints convert microscopic stacking disorder into persistent macroscopic order, providing a general framework for understanding and exploiting rapid solidification processes far from equilibrium.

**Cryo membrane liquid cell and homogeneous nucleation**

Micron-scale water droplets were prepared by encapsulating a small volume of liquid water (~1 μL) between two continuous amorphous carbon membranes (~20 nm thick), forming a membrane liquid cell suitable for cryogenic transmission electron microscopy (TEM) (Fig. 1A). High angle annular-dark-field (HAADF) imaging confirms that this approach reproducibly yields isolated water droplets with diameters ranging from ~1 to 10 μm (Fig. 1B; Figs. S1a–c). Droplets smaller than ~5 μm are predominantly round, whereas larger droplets often exhibit irregular outlines (Figs.

S1b and S1d). Energy-filtered TEM measurements reveal a thin concave lens-like thickness profile, with thickness decreasing continuously from the center toward the edge (Figs. S1f–h). For example, a representative 2 μm-diameter droplet exhibits a central thickness of ~33 nm (Fig. 1C), enabling molecular-scale TEM imaging of the nucleation center after freezing.

The temperature of the liquid-nitrogen-cooled TEM holder can be precisely controlled from −170 °C to 100 °C. During temperature cycling, water droplets reproducibly melt at 0 °C and freeze at −40 (±0.5) °C (Fig. S2a), allowing repeated water–ice transformation loops. Despite confinement by amorphous carbon membranes, crystallization proceeds via homogeneous nucleation, supported by two observations. First, the freezing temperature is independent of droplet size and coincides with the known homogeneous nucleation temperature of deeply supercooled water (−40 °C) [37]. Second, nucleation sites are spatially random. For example, when the same droplet undergoes multiple freezing–melting cycles, nucleation initiates at different locations each time (Fig. S2b). Statistical analysis of nucleation centers from 100 droplets, normalized by droplet diameter, shows that the nucleation probability scales with droplet volume rather than surface area (Figs. S2c and S2d) [38], ruling out active surface-mediated nucleation sites previously proposed [39].

**Globally ordered octahedral dendritic morphology**

Upon freezing at −40 °C, deeply supercooled water droplets crystallize rapidly (Fig. S3), producing ice crystals with a strikingly regular growth morphology (Fig. 1D–G). When imaged along the $[110]_c$ or $[100]_h$ zone axis (c and h stand for cubic and hexagonal, respectively), the crystals display a single nucleation core and eight dendritic branches radiating at fixed angles. Collectively, these branches form an octahedral, globally cubic-symmetric pattern, reminiscent of a Maltese cross (Fig. 1E) but distinct from the sixfold symmetry typical of hexagonal snowflakes formed near equilibrium. This octahedral dendritic pattern is robust, appearing in both small ice crystals with nearly circular outlines (Fig. 1D) and larger crystals with irregular external contours (Fig. 1F).

To minimize electron-beam damage associated with high-probe-current scanning TEM (STEM) imaging, we employed central-beam-blocked dark-field (CBBDF) TEM imaging (Fig. S4) at ultralow electron dose rates (below 0.5 $e^-/nm^2s$) to visualize stacking faults distributions [40]. The CBBDF image (Fig. 1G) reveals a high density of stacking faults throughout the dendritic branches. High-resolution TEM (HRTEM) imaging further shows that the branches do not emanate from a geometric point. Instead, growth originates from a nanoscale core of stacking-disordered ice ($I_{sd}$) with an average diameter of 90 ± 20 nm, as indicated by the green dashed circle in Fig. 1G. These observations demonstrate that a globally ordered morphology emerges reproducibly from a deeply nonequilibrium freezing process, despite pronounced local structural disorder.

**Cubic twinning and global symmetry**

Based on stacking-fault orientation and crystallographic relationships, the eight dendritic branches can be classified into primary, secondary, and tertiary groups (Fig. 1G). The primary dendrites share the same stacking-fault plane as the nucleation core and represent direct extensions of the core lattice along opposite basal directions. The basal planes in secondary dendrites are rotated by 109.5° relative to the primary branches, indicating a cubic twinning relationship. Tertiary dendrites complete the eight-branch motif but do not exhibit visible stacking-fault contrast under the present imaging conditions.

To determine the crystallographic orientation relationship among all branches and to access whether tertiary dendrites are $I_{sd}$, we employed three-dimensional electron diffraction (3D-ED) [41], schematically illustrated in Fig. 1H. This technique reconstructs reciprocal-space diffraction volumes and enables virtual rotation of the crystal over 360°, allowing access to arbitrary zone axes even when physical specimen tilting is limited. The reconstructed diffraction patterns show that all eight branches, including the tertiary dendrites, consist of $I_{sd}$ rather than pure $I_h$ or $I_c$. In each branch, the basal plane is aligned with one of the eight {111}c planes corresponding to the octahedral facets of a cubic lattice (Fig. 1I). The projected diffraction patterns along $[110]_c$ (Fig. 1J and Fig. S5) acquired from tertiary dendrites exhibit characteristic streaking associated with a high density of stacking faults, confirming their $I_{sd}$ nature.

3D-ED pattern (Fig. 1K) from pairs of primary and tertiary dendrites further reveals that the crystallographic orientation between any two branches is fixed at 109.5°, confirming that all branches are related by twinning on $\{111\}_c$ planes. Collectively, the eight dendrites occupy the eight octahedral facets of a cubic lattice, giving rise to a globally cubic-symmetric morphology despite the strong stacking disorder within each individual branch. When diffraction from all branches within a single ice crystal is combined, the resulting 3D-ED pattern resembles that of a simple cubic system, exhibiting fourfold symmetry along the [100] direction (Fig. 1L) and threefold symmetry along the [111] direction (Fig. S6). Thus, although each dendrite is highly stacking-disordered, the ice crystal as a whole displays robust global cubic symmetry.

**Stacking disorder and cubicity evolution**

The two-dimensional confinement of the membrane liquid cell enables precise localization of the nucleation center with molecular-resolution imaging. Using low-dose molecular-resolution HRTEM, we resolved the stacking sequence and quantified cubicity following the procedure described in Fig. S7 [42,43]. Low-dose HRTEM imaging of the nucleation core (Fig. 2A) reveals a strongly stacking-disordered structure dominated by hexagonal ice $I_h$, with isolated monolayers of cubic ice $I_c$ as stacking faults. Quantitative analysis of the stacking sequence yields a low initial cubicity of ~0.35 (Fig. 2B), indicating that cubic ice is present only as single layers at the earliest stage of growth. This observation contrasts with several molecular dynamics simulations that predict a cubic-rich nucleus during early crystallization [24,26].

As crystallization progresses outward from the nucleation core, the stacking order evolves systematically. In secondary dendrites (Fig. 2C), hexagonal and cubic segments exhibit comparable thicknesses, corresponding to an increased cubicity of ~0.53 (Fig. 2D). Notably, in the primary dendrites, cubicity rises progressively from ~0.35 at the nucleation center to ~0.53 near the primary–secondary branching interface, reaching values similar to those of the secondary dendrites. In tertiary dendrites (Fig. 2E), cubic segments become dominant, while hexagonal layers occur primarily as stacking faults, yielding a further increase in cubicity to ~0.62 (Fig. 2F). In parallel with this increase in cubicity, the thickness of individual cubic segments grows from isolated monolayers near the core to continuous regions approaching ~1 nm in thickness at larger growth radii (Fig. S7d-f), consistent with previous reports using molecular dynamics simulations [25]. This monotonic increase in both cubicity and cubic-segment thickness demonstrates that stacking order in ice formed from deeply supercooled water evolves dynamically during growth and is governed by kinetic effects, rather than being fixed at the point of nucleation.

**Nanoscale cubic ice germs as kinetic branching seeds**

The formation of the octahedral dendritic pattern can be understood as a multistage kinetic growth process mediated by nanoscale cubic ice germs, highlighted by the yellow parallelogram in Fig. 3A. Initially, the $I_{sd}$ nucleation core (#0) grows anisotropically, with faster growth along basal planes and slower growth on prismatic planes, giving rise to two primary dendrites (#1 and #2) (Fig. 3B). This initial stage continues until a cubic ice germ forms on the basal surface of the $I_{sd}$ core. Acting as a secondary nucleation site, this cubic ice germ seeds the formation of secondary dendrites (#3 and #4) that obey the twin law of cubic ice. High magnification image shown in Fig. 3C confirms that the lattice structure in the marked cubic ice germ is a pure, defect free $I_c$, as confirmed by HRTEM image simulations (Fig. S8). By locally imposing cubic twinning symmetry (Fig. 3C), these nanoscale cubic germs define the branching orientation at 70.5° (or the complementary 109.5°) and thereby govern the development of the octahedral growth morphology. Because secondary branching occurs exclusively on basal planes rather than prismatic surfaces, the primary dendrites (#1 and #2) are consistently connected to each other through the core (#0), whereas the secondary dendrites (#3 and #4) never connect with one another (Fig. 1G and Fig. 3B).

Subsequent growth of the primary and secondary dendrites generates new grain boundaries that intersect their respective basal planes at an equal angle of ~35° (half of 70.5°), indicating comparable growth rates perpendicular to the basal planes. These grain boundaries provide favorable sites for the nucleation of additional cubic ice germs (Fig. 3D), which subsequently seed the formation of tertiary dendrites. High-resolution TEM directly resolves these nanoscale cubic ice germs are defect free at branching junctions (Fig. 3B–D), confirming their role as symmetry-enforcing kinetic seeds. Statistical analysis of cubic germ sizes shows that the cubic germs responsible for the nucleation of secondary dendrites (ranging from 3 to 13 nm) are smaller than those associated with tertiary dendrites (around 20 nm), as supported in Fig. S9. Each newly formed cubic germ induces additional twinning events, doubling the number of branches and ultimately producing eight dendrites aligned with the $\{111\}_c$ facets of a cubic lattice (Fig. 1I). Through this hierarchical branching process, the growing ice progressively fills the droplet volume and collectively adopts an overall cubic symmetry, as confirmed by 3D electron diffraction (Fig. 1L and Fig. S5). This growth mechanism is further supported by molecular dynamics simulations with the coarse-grained Bond Order Potential model (Fig. 3E and Fig. S10) [28], which independently reproduce the preferential nucleation of cubic ice on basal planes of $I_{sd}$ during rapid growth.

**Late-stage thermodynamic growth of hexagonal ice**

In larger ice crystals, regions of pure hexagonal ice $I_h$ emerge at distances exceeding ~1-1.5 μm from the nucleation center. These regions are identified using scanning electron nanobeam diffraction (SEND) combined with unsupervised machine-learning clustering, which distinguishes Ice $I_h$ by the absence of stacking-fault streaking in the diffraction patterns (Fig. 4A–E). In contrast, smaller crystals remain predominantly stacking-disordered, indicating that pure Ice $I_h$ forms only at later growth stages.

SEND, with a beam size of ~ 250 nm, was applied to a ~6.5 μm ice crystal frozen from deeply supercooled water and exhibiting eight radiating dendrites in the HAADF image (Fig. 4A). Diffraction patterns acquired across the crystal were classified using unsupervised machine learning [44], with silhouette-score analysis [45] indicating optimal clustering at k = 7 (local maximum 0.273) and k = 13 (global maximum 0.302). At k = 7, the crystal is segmented into regions corresponding to six twinned dendrites with distinct orientations, while at k = 13 these regions are further subdivided. Summed diffraction patterns reveal that most areas retain a stacking-disordered structure, whereas selected outer regions correspond to pure hexagonal ice $I_h$, confirmed by the absence of streaking and stacking faults in both diffraction patterns and high-magnification images (Fig. 4D–E).

The emergence of pure $I_h$ in these outer regions is attributed to the exothermic nature of freezing [46,47]. Latent heat released during crystallization progressively raises the temperature at the advancing growth front as it propagates away from the nucleation center. Three-dimensional numerical simulations of droplet freezing (Fig. 4F and Fig. S11) show that, for a 6 μm-diameter droplet, the front temperature increases by ~10 °C after ~2 μm of propagation, entering a thermodynamic regime favorable for hexagonal ice formation. In contrast, smaller droplets (~1 μm diameter) experience only a modest temperature rise (~3–4 °C), insufficient to overcome kinetic constraints, explaining the absence of pure $I_h$ in small crystals (e.g., Fig. 1D and Fig. 1G).

**Conclusions:**

As schematically summarized in Fig. 5, ice formed from deeply supercooled water follows a previously unrecognized nonequilibrium crystallization pathway governed by the gradual thickening of cubic ice. Homogeneous nucleation first produces stacking-disordered ice containing isolated monolayer cubic layers. The progressive thickening of these cubic layers constitutes the central kinetic mechanism that controls the entire morphology. Once a critical thickness is reached, nanoscale, defect-free cubic ice germs emerge on the basal planes of the disordered lattice and act as facet-registered kinetic seeds that enforce cubic twinning and drive the sequential multiplication of growth branches. This thickening-driven pathway reproducibly generates robust eight-branched dendrites with global cubic (octahedral) symmetry, in which the “core” is not a single point but a connected network of primary branches that defines the overall symmetry.

At later stages, latent heat release induces a crossover to the thermodynamically favored hexagonal phase; remarkably, this transformation does not erase the symmetry established during the kinetically dominated growth stage. Together, these observations reveal a transition from a kinetically controlled, stacking-disordered growth regime near the nucleation center to a thermodynamically controlled hexagonal growth regime at larger distances. This work establishes a direct connection between atomic-scale disorder and emergent macroscopic symmetry, demonstrating how extreme kinetic driving forces can generate higher-order order far from equilibrium and providing a general framework for kinetic pathway engineering in rapid crystallization processes.

## Materials and Methods

**Water and ice droplets Formation**: Water droplets are formed by sandwiching a drop of liquid water (~1 ml) between two transmission electron microscopy (TEM) grids with continuous amorphous carbon films (15 - 20 nm in thickness). The sandwich is loaded into a Gatan liquid nitrogen cooling holder before water evaporates. Before loading into the TEM chamber, the liquid cell was pre-pumped in a vacuum pumping station to remove excess water. More importantly, this process forces the two membranes into close contact, effectively sealing the remaining water within the cell. Although the microscope column operates under high vacuum, the liquid confined inside the sealed cell remains in a near-ambient-pressure environment. The temperature at the TEM sample area can be set to any number from -170 °C to 100 °C, such that the ice crystallization and melting process can be in situ studied. Water droplets obtained from this method are between 1 to 10 μm in diameter, 30 nm to 150 nm in thickness. The geometry of water trapped between two films looks like a thin convex round lens. Before loading TEM grids into a TEM chamber, the cold finger in TEM is cooled with liquid nitrogen for at least one hour to ensure there is extra ice condensation on surfaces of the TEM grids.

**TEM Characterization:** Transmission electron microscopy is carried out using the Argonne Chromatic Aberration-corrected TEM (ACAT), a FEI Titan 80–300 transmission electron microscope equipped with an image corrector to correct both spherical and chromatic aberrations. Thickness mapping is carried out using energy-filtered TEM thickness mapping technique with a Gatan image filter system. EELS is acquired using the ACAT operating at 200 kV. Energy dispersive spectroscopy (EDS) mapping and 3D electron diffraction is carried out using a FEI Talos F200X scanning TEM (STEM).

**Low-dose Low-magnification TEM**: The typical electron dose rate of low magnification TEM imaging and selected-area electron diffraction (SAED) is below 10 $e^- Å^{-2}s^{-1}$. A total dose of below 100 $e^- Å^{-2}$ enables low-magnification imaging and SAED on the same area for 10 minutes without any noticeable change in ice. In addition, a direct electron detection camera Gatan K2 with better detection efficiency than CCD is used for recording images. Typically, 20 frames with exposure time of 0.1 second are recorded for each image. The electron beam is always blanked before opening K2 shutter. 20 frame images are aligned with drift correction and summed for better SNR.

**Low-dose high resolution transmission electron microscopy (HRTEM)**: Low-dose HRTEM images are taken based on the following procedures: 1) After searching the suitable ice nanocrystal with a weak electron beam at low-magnification and the location is recorded. 2) Move to a carbon film away from the region of interest (ROI) and do all necessary TEM alignment including aberration correction alignment. After the alignment is done, first electron beam is blocked, and then set up an adequate value of electron beam intensity such as dose-rate below $5\times10^2$ $e^-$ $Å^{-2}s^{-1}$ for HRTEM images. 3) Recall the ROI and increase the magnification for HRTEM imaging. Start taking images first using the direct electron detection camera with an exposure time of 0.1 seconds and then open the beam blank to exposure electron beam on the ROI for recording a set of 50 images. Only the first few frames without noticeable electron beam damage (total dose below $1\times10^4$ $e^-$ $Å^{-2}$) are chosen for further imaging analysis.

To avoid any influence of electron irradiation on ice crystallization from supercooled water, the electron beam was blanked during the supercooling process, except for the high-speed imaging experiments shown in Fig. S3. Low-dose high-resolution transmission electron microscopy was performed only after the crystallization process was completed. Therefore, the crystallization occurred in the absence of continuous electron-beam exposure.

**HRTEM image and diffraction patterns simulation:** MacTempas and CrystalKit softwares are used for HRTEM image and diffraction pattern simulations.

**Central-beam-blocked dark-field (CBBDF) TEM imaging**: A custom-made objective aperture is fabricated to allow passage of all diffracted electron beams except the transmitted beam as shown in Supplementary Information Fig. S3. CBBDF TEM uses a board electron beam, which enables imaging ice at a much lower dose rate below 0.5 $e^-/nm^2s$, providing Z-contrast like HAADF and diffraction contrast for imaging stacking faults.

**3D Electron Diffraction:** The 3-dimensional continuous-rotation electron diffraction (3D-cRED) datasets are acquired by stepwise tilting the sample with a collection angle of −45° to 45° and 400 frames in total. The reciprocal lattice pattern is then reconstructed using a Python script and the RED processing software package developed by Wan et al.[48]. The script can be downloaded from https://github.com/danielzt12/EDiff_3D_projection/.

**HRTEM image phase search:** First, apply the image processing to locate atomic columns from the HRTEM image. Then using the local symmetry, we can precisely determine the stacking sequence from a HRTEM image (Fig. S7).

**MD simulations details:** The details are described in Supplementary Information Fig. S10

**Unsupervised machine learning method for ice classification:** 40x40 NBD patterns on an entire ice droplet are acquired using a CCD camera with a beam size of 50 nm and step size of 250 nm. Electron diffraction patterns are processed using an unsupervised machine learning method[49]. The raw diffraction patterns (40x40 NBD patterns) are aligned using the central beam position before being cropped to the size of 600x600 pixels. Subsequent processing included binning the cropped images to 120x120 pixels, thresholding and rescaling to logscale. Sihouette score suggests two optimized number of clusters[50], a local maxima (0.273) at k=7 and a global maxima (0.302) at k=13. With k=7, the real space image is divided into 7 colored areas as shown in Fig. 4b. Those areas (except for the empty space) correspond to 6 twinned grains with different orientations. With k=13, some of the areas are further divided into small areas with different hatch patterns. Among these patterns, three correspond to tertiary dendrites without any diffraction streaking and four correspond to the diffraction patterns of ice $I_{sd}$ along a zone axis equivalent to the $[110]_c$ or $[100]_h$ in primary and secondary dendrites. In addition to these ice structures, the machine learning method finds three areas that can be indexed as ice $I_h$. As shown Fig. 4b, regions with ice $I_h$ structures are observed when the diameter is 1-3 μm away from the nucleation center.

**Temperature evolutions during freezing of supercooled water droplets**: The numerical simulation of temperature evolution during freezing of supercooled micro-sized water droplet are conducted using the approaches developed by Meng and Zhang[46,47]. (Supplementary Information Fig. S11)

**Acknowledgments:**

**Funding**: Work performed at the Center for Nanoscale Materials, a U.S. Department of Energy Office of Science User Facility, was supported by the U.S. DOE, Office of Basic Energy Sciences, under Contract No. DE-AC02-06CH11357. P.Z. thanks the financial support from National Natural Science Foundation of China (52576220) and Natural Science Foundation of Shanghai Municipality (25ZR1401209). We thank helpful discussion with Prof. Valeria Molinero, Dr. Zhaonan Meng, and Mr. Neel Koritala.

**Author contributions:** Y.L. and J.W. conceived the project and designed the experiments. Y.L. and J.W. conducted the TEM experiments. Y.L. and T.Z. performed ice phase identification from HRTEM images. Y.L., L.Y., and T.Z. performed 3D electron diffraction. Y.L. M.S. J.W. carried out diffraction and HRTEM image simulations. Y.L., T.E.G. and J.W. performed high-speed imaging experiments. T.Z. performed machine learning unguided phase search. W.G. and P.Z. performed numerical simulation of temperature evolution during freezing of supercooled micro-sized water droplets. S.B. and S.S. carried out MD simulations. Y.L. T.Z. and J.W. wrote the manuscript. All authors discussed the results and commented on the manuscript.

**Competing interests:** Authors declare no competing interests.

**Data and materials availability:** All data in the main text or the supplementary materials are available from the corresponding author upon request.

Correspondence and requests for materials should be addressed to tzhou@anl.gov; ssankaranarayanan@anl.gov; zhangp@sjtu.edu.cn; jwen@anl.gov.

**List of Supplementary Materials**:

Figs. S1 to S11

References (1-47)

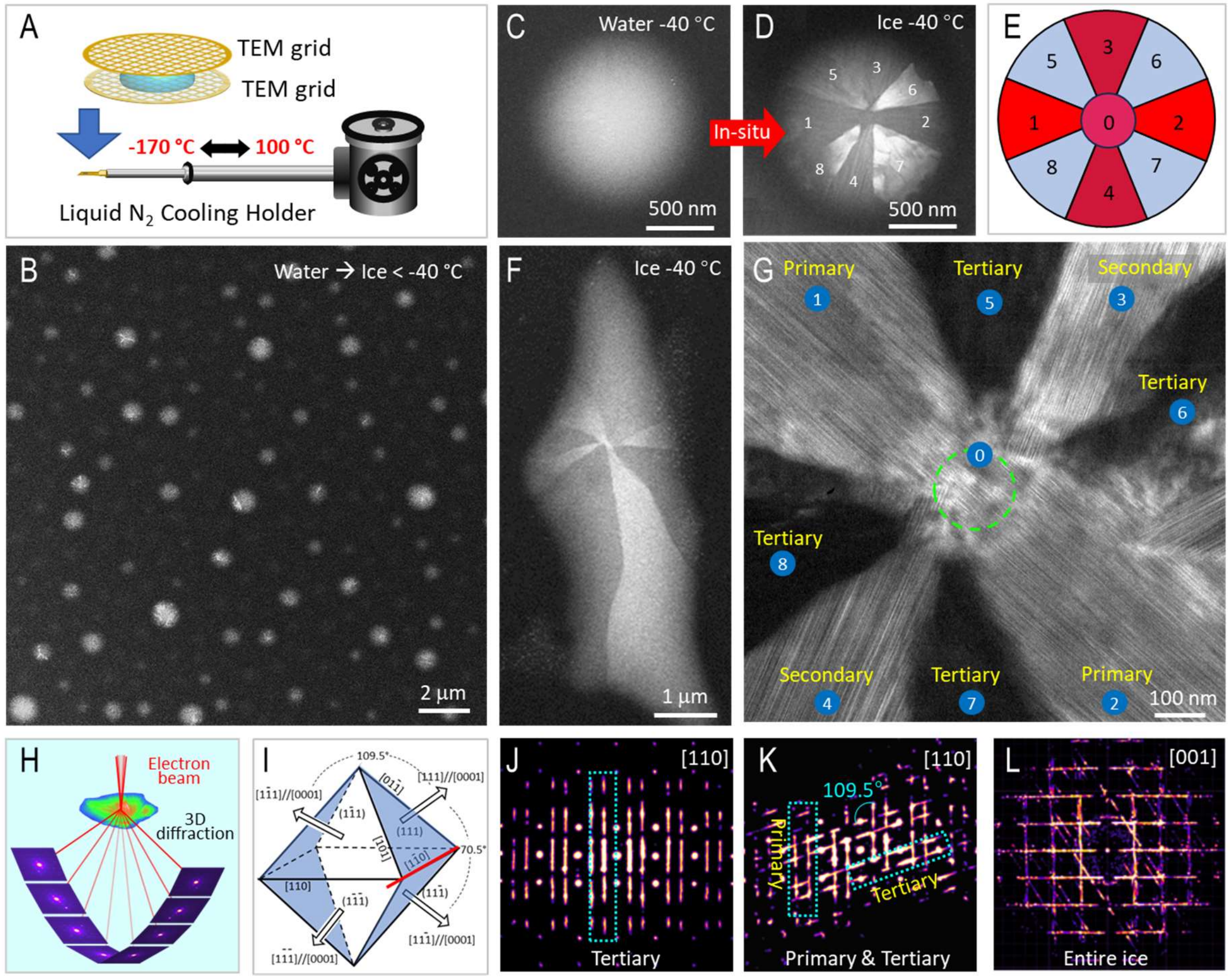

**Fig. 1 Maltese cross pattern in ice frozen from supercooled water at -40°C**. (A) Schematically showing the formation of water droplets formed between TEM grids. The sandwich is loaded in a liquid $N_2$ cooling TEM holder for in-situ study of water/ice transformation. (B) Micro-sized water droplets confined between two TEM grids coated with continuous amorphous carbon films freeze into ice when the temperature is lowered below −40 °C. (C) Micro-sized supercooled water just above -40 °C and suddenly turns into 8-branch dendritic ice (~1 μm) in (D). (E) A Maltese cross pattern illustrating the typical characteristics of ice formed from supercooled water includes a nucleation core (#0), two primary dendrites (#1, #2) extending directly from the core along the prism surfaces, two secondary dendrites (#3, #4) not directly connected to each other, and four tertiary dendrites (#5-#8). (F) A ~10 μm ice droplet also shows the 8-branch dendritic morphology. (G) A CBBDF TEM image of 8-branch dendritic ice showing the features summarized in the schematic in (E). Note primary and secondary dendrites are twinned with an angle of 109.5°

with a central symmetry. (H) Schematic of 3D electron diffraction technique. (I) Each basal plane of eight twinned dendrites is on one of octahedral facets. (J) Projected 3D diffraction pattern from a tertiary dendrite showing it is an ice $I_{sd}$ as two extra rows of diffraction streaking due to SFs are observed. (K) Projected 3D diffraction pattern from primary and tertiary dendrites showing the twinning angle of 109.5°. (L) Projected 3D diffraction pattern from eight dendrites along $[001]_c$.

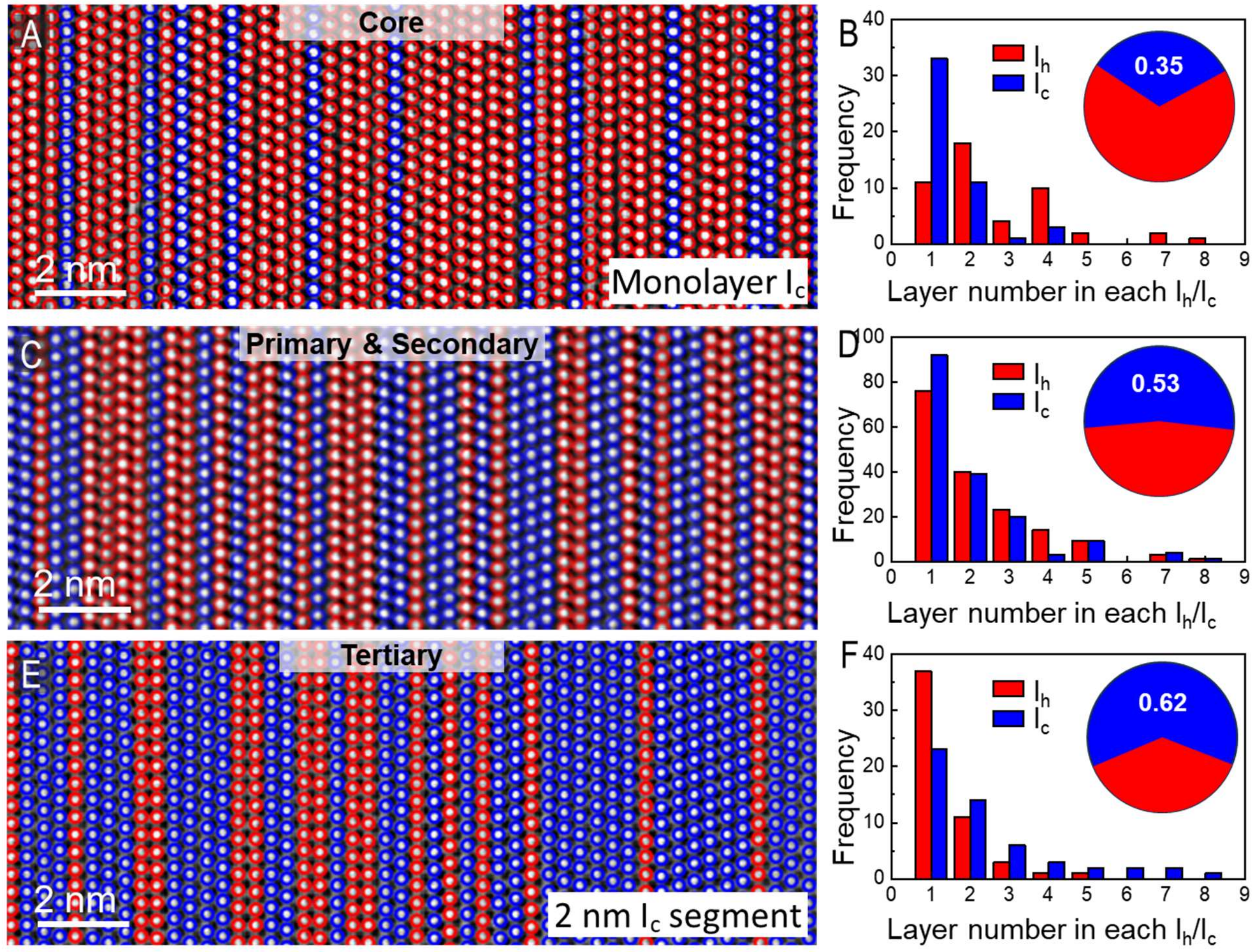

**Fig. 2 Stacking disorder and cubicity evolution.** (A) low-dose HRTEM image of a typical ice $I_{sd}$ at the core with a cubicity of 0.35 (B). Many isolated single layers of ice $I_c$ (bule) appear as SFs in the ice $I_h$ (red) segments as shown from the layer number distribution of each ice $I_c$ or $I_h$ segments. (C) The ice $I_{sd}$ in the secondary dendrites have a cubicity of 0.53 (D). The probability of monolayer ice $I_h$ (red) and ice $I_c$ (blue) in the regions is almost equal. Within the primary dendrites, the cubicity increases continuously from the core to primary–secondary interface. (E) The ice $I_{sd}$ in the tertiary dendrites exhibits a cubicity of 0.62 (F). Cubic ice $I_c$ (blue) segments significantly thicker, reaching ~ 2nm. Large area HRTEM images and definition of $I_h$ and $I_c$ are shown in Fig. S7.

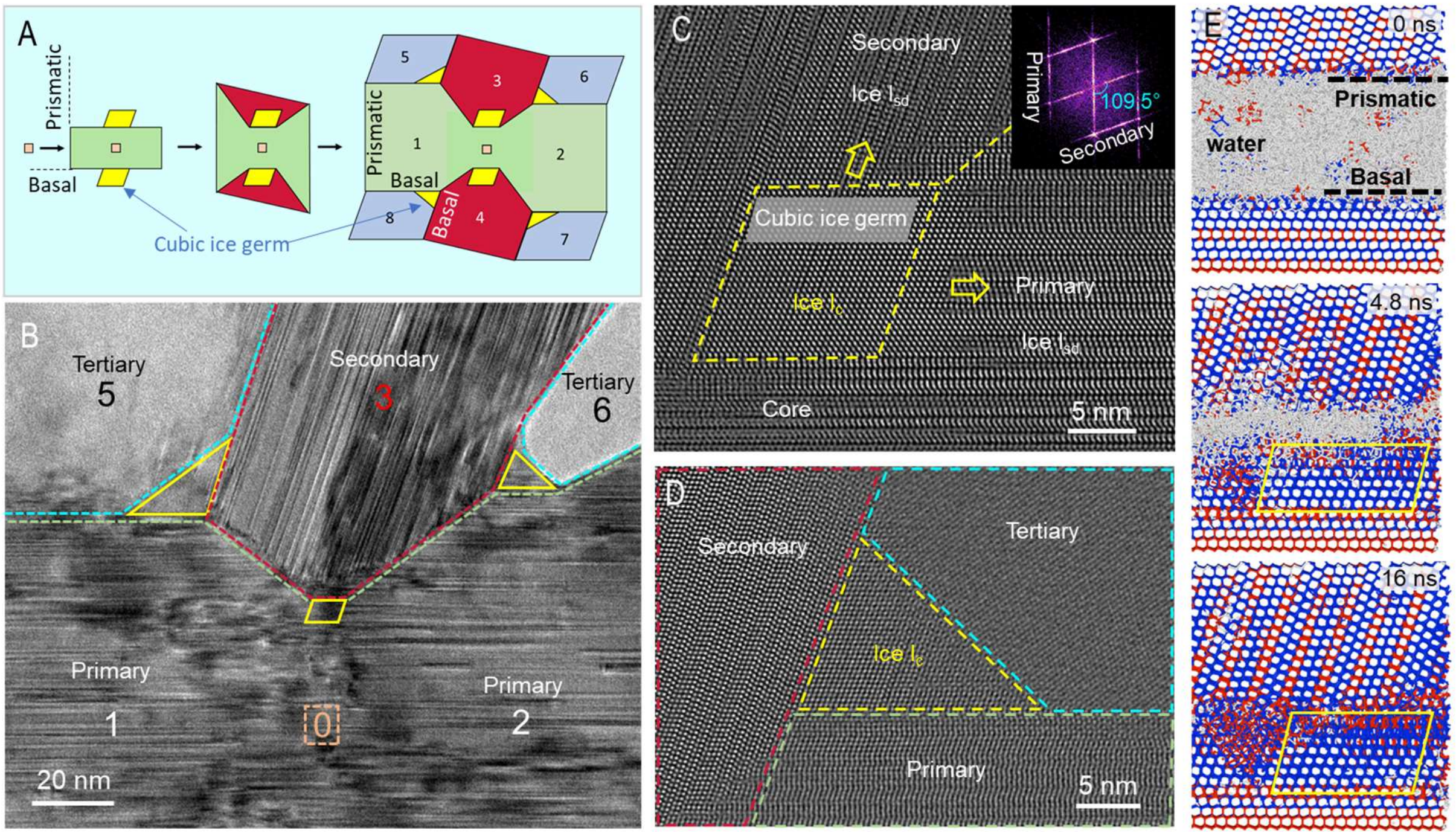

**Fig. 3 Globally ordered octahedral dendritic morphology govern by nanoscale cubic ice germ**. (A) Schematic showing the regular growth pattern: primary dendrites extending from prism surfaces of the ice $I_{sd}$ core and secondary and tertiary dendrites branch out from a nanoscale cubic ice germ nucleated on the basal plane of ice $I_{sd}$. (B) A high-magnification TEM image of eight-branch dendritic ice showing the distribution of the core, primary, secondary, and tertiary dendrites. (C) Low-dose HRTEM image from the nucleation point of secondary dendrite (the small yellow parallelogram in (B)), showing a cubic ice $I_c$ germ on the basal plane at the tip of the secondary dendrite. The fast Fourier transform (FFT) pattern (the inset) shows two streaking lines with an angle of 109.5°, indicating the primary and secondary dendrites are in a cubic twinning relationship. (D) Low dose HRTEM image from the tertiary dendrite, showing a nanoscale cubic ice germ highlighted by the yellow triangle. (E) Snapshots of an MD simulation illustrating the extension growth along the prism surfaces (upper row), the nucleation and growth of a cubic ice germ on the basal plane (highlighted in yellow) during the formation of twin boundaries in $I_{sd}$ from a liquid in dendritic morphology. The bottom $I_{sd}$ aligns its basal plane with the liquid, while the tilted top $I_{sd}$ exposes prismatic plane to the liquid at 230K.

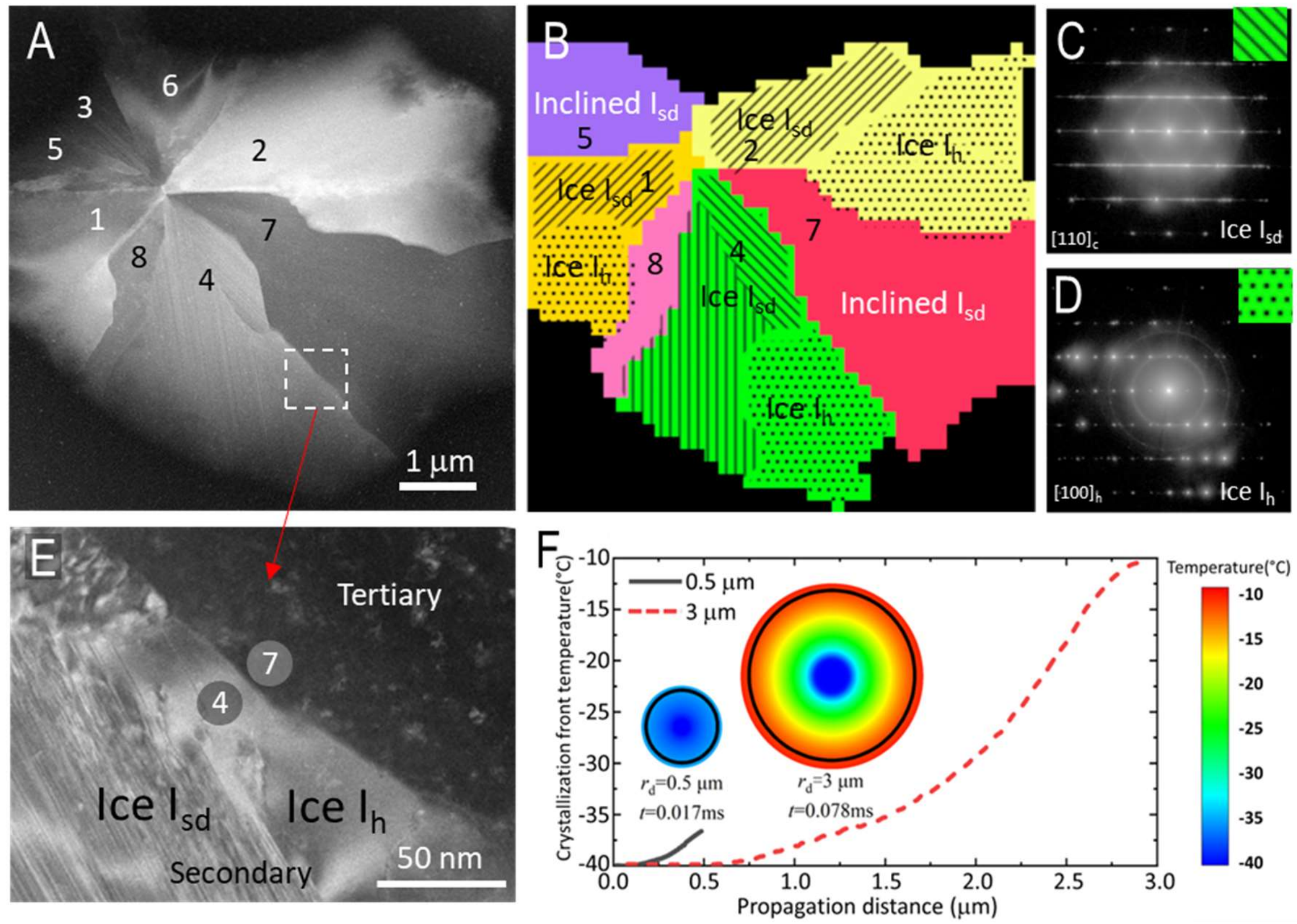

**Fig. 4 Late-stage thermodynamic growth of hexagonal ice due to latent heat release.** (A) HAADF image of ice (radius > 5 µm) used for SEND mapping. (B) AI-assisted classification of ice structures, revealing the emergence of hexagonal ice ($I_h$) along the sides of dendrites. (C,D) Representative diffraction patterns from regions identified as stacking-disordered ice ($I_{sd}$) and hexagonal ice ($I_h$), respectively. (E) Magnified view of the boxed region in (A), showing the disappearance of stacking faults within the $I_h$ domain. (F) Crystallization-front temperature as a function of propagation distance from the nucleation center for supercooled water droplets with radii of 0.5 and 3 µm. A pronounced temperature rise is evident in the temperature map during freezing (inset) for the 3 µm droplet; the black circle marks the crystallization front.

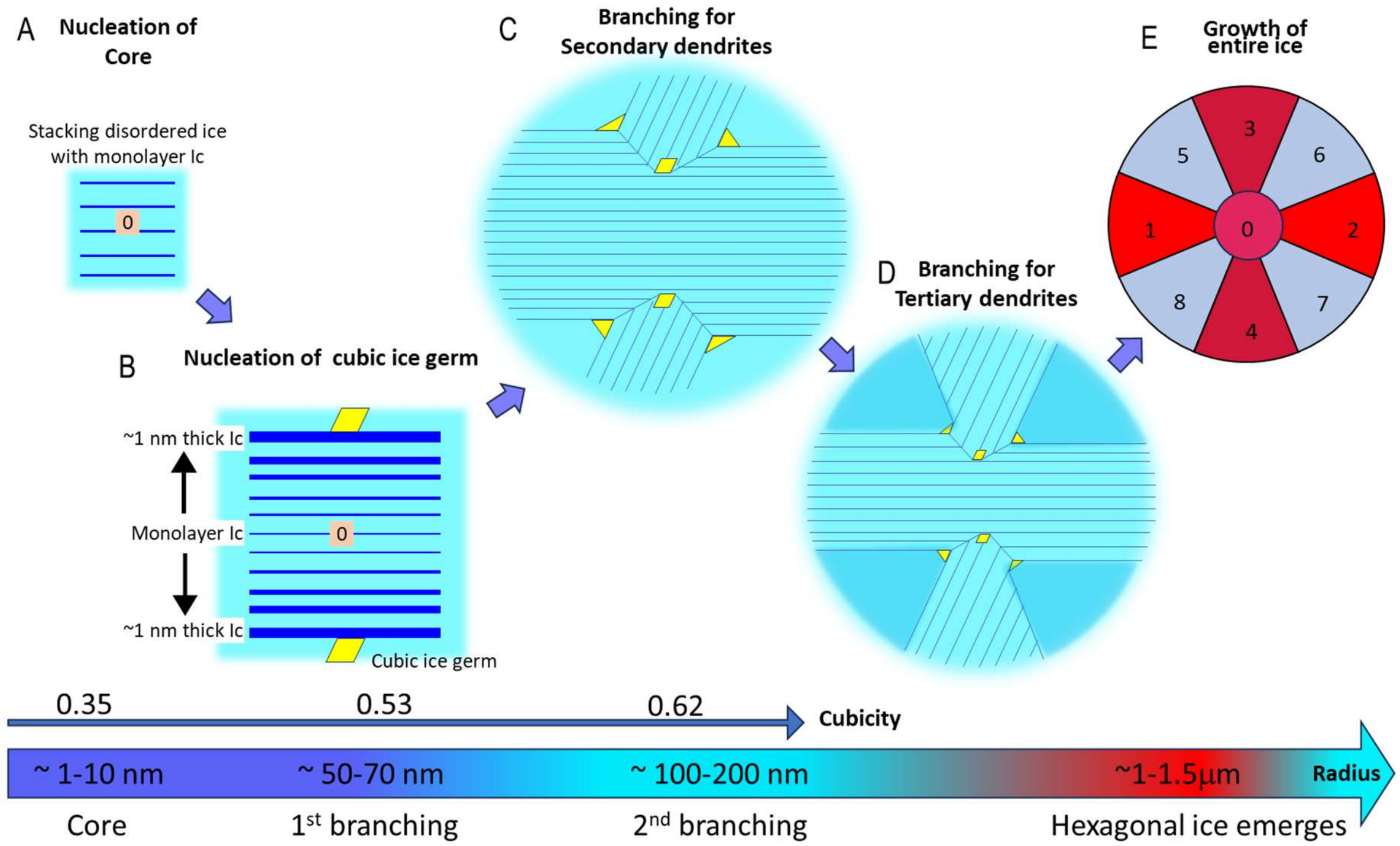

**Fig. 5 Freezing dynamics of deeply supercooled water.** (A) Homogeneous nucleation yields stacking-disordered ice ($I_{sd}$) composed of isolated monolayer cubic ice ($I_c$) layers. (B) During rapid growth, cubic components thicken and nanoscale cubic ice germs emerge. (C,D) These cubic germs kinetically promote twinning, leading to the development of an octahedral dendritic morphology that exhibits global cubic symmetry, shown in (E). At later stages, latent heat release induces a crossover to thermodynamically favored hexagonal ice growth. Representative length scales and cubicities for each stage are indicated at the bottom.

# Supplementary Information

## From Stacking Disorder to Cubic Order: Ice Crystallization from Deeply Supercooled Water

Yulin Lin[1,2], Weimin Guo[3], Suvo Banik[1], Tao Zhou[1*], Thomas E. Gage[1], Lei Yu[1], Maksim A. Sultanov[1], Martin Holt[1], Subramanian Sankaranarayanan[1*], Peng Zhang[3*], Ilke Arslan[1], Jianguo Wen[1*]

**Affiliations:**

[1]Center for Nanoscale Materials, Argonne National Laboratory, Lemont, Illinois, United States

[2]Institute for Advanced Studies, College of Chemistry and Molecular Sciences, Wuhan University, Wuhan, P. R. China

[3]Institute of Refrigeration and Cryogenics, Shanghai Jiao Tong University, Shanghai, P.R. China

[*]Corresponding authors: Tao Zhou (tzhou@anl.gov); Subramanian Sankaranarayanan (ssankaranarayanan@anl.gov); Peng Zhang (zhangp@sjtu.edu.cn); Jianguo Wen (jwen@anl.gov)

**List of Supplementary Materials**:

Figs. S1 to S11

References

## Supplementary Information Fig. S1

### Water droplets dimensions

The diameters of water droplets are directly measured from HAADF images. Thickness mapping of a water droplet was acquired using energy filtered TEM (EFTEM). The thickness of the sample is measured in units of mean free path λ for inelastic scattering followed by: $t/\lambda = \ln(I_{unfilt}/I_0)$, where $I_{unfilt}$ is the total unfiltered image intensity, $I_0$ is the zero-loss intensity [1], and λ is total mean free path λ for inelastic scattering. For an accelerating voltage of 200 kV, the inelastic mean free path for water (130 nm) is very close to that of carbon film (107 nm). Statistical analysis reveals a linear correlation between droplet diameter and the thickness at their centers.

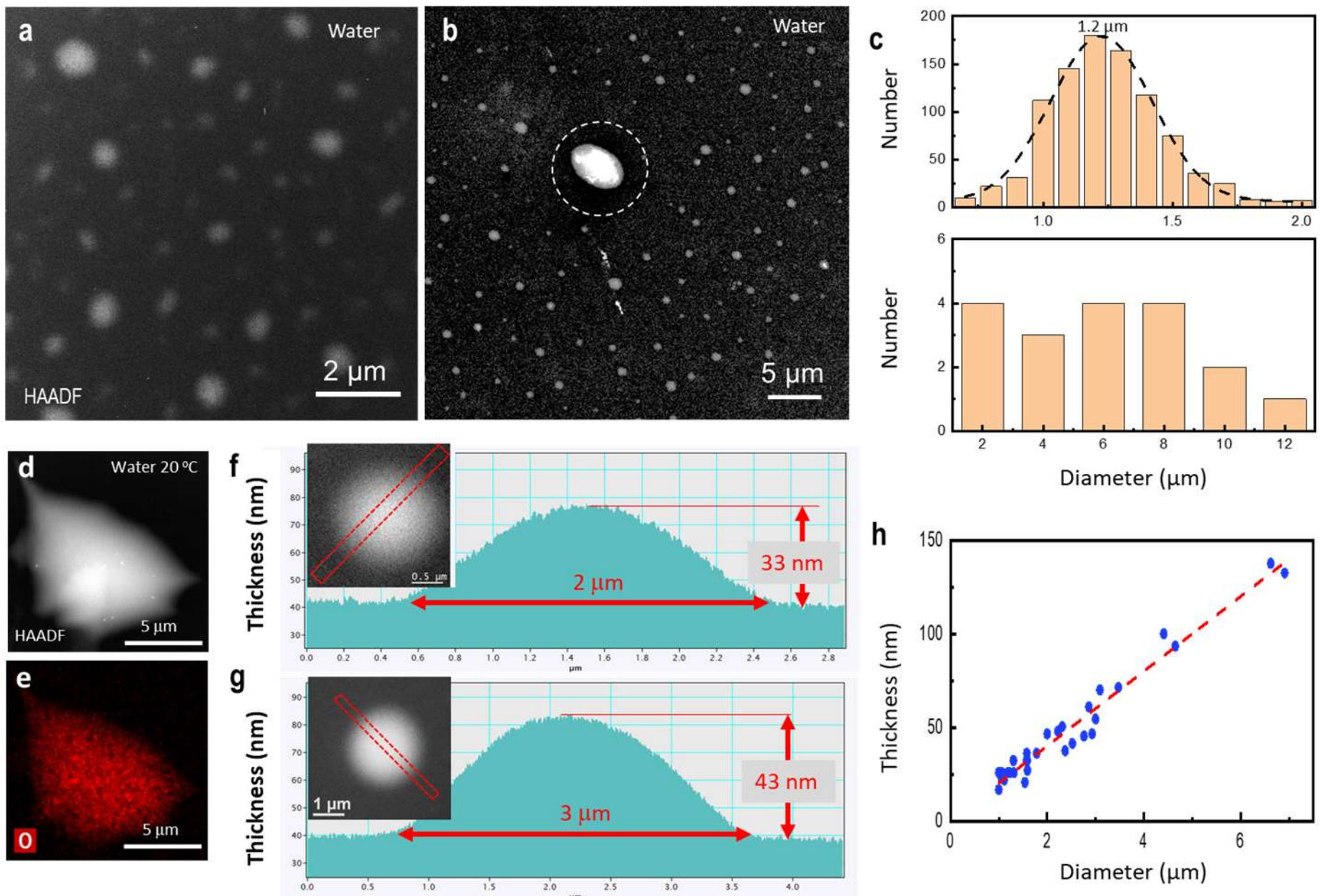

**Fig. S1 Dimensions of water droplets**. HAADF image showing the water droplets a) < 2 μm; b) > 2 μm trapped between two amorphous carbon films. c) Statistical analysis shows most water droplets are 1-2 μm in diameter and some of them are larger than 5 μm. d) HAADF image and e) EDS mapping showing oxygen only for a water droplet larger than 5 μm. f) EFTEM thickness mapping and a thickness profile of a 2 μm water droplet. g) EFTEM thickness mapping and a thickness profile of a 3 μm water droplet. h) The dependence of maximum thickness with water droplet diameter.

## Supplementary Information Fig. S2

### Randomized Nucleation during Repeated Water–Ice Phase Cycling

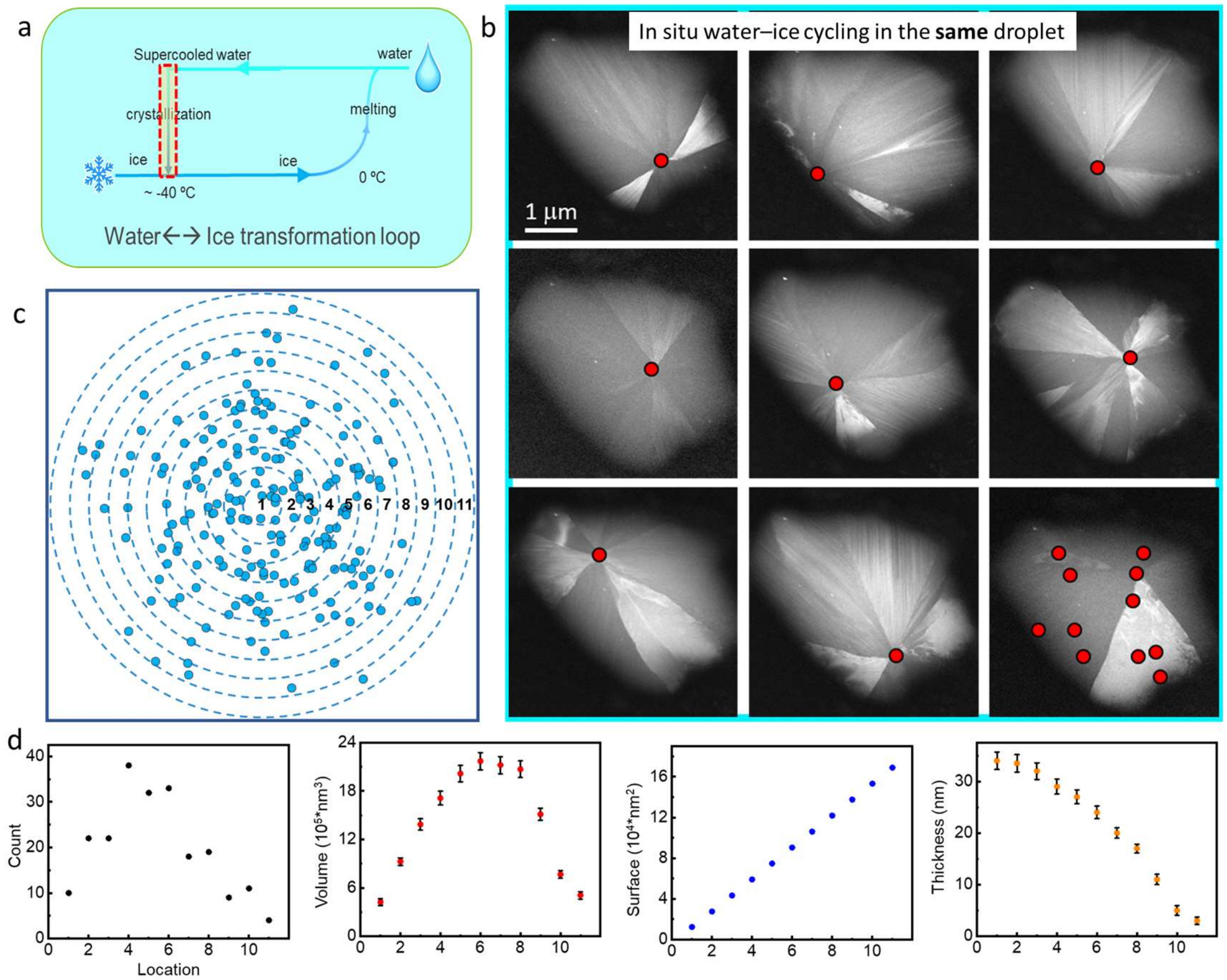

**Fig. S2 Random nucleation centers in supercooled water droplets.** a) Micro-sized water droplets go through a freezing/melting loop. b) HAADF images of ice formed from the same water droplet after multiple temperature cycles. At each cycle, the sample was heated to 5 °C and then cooled down to -50 °C at a cooling rate of -0.3 °C/s, since we find that micro-sized water freezes at -40 °C but melts at 0 °C. Nucleation centers are marked in the last figure. The 8-branch dendritic feature of ice nanocrystals is not obvious due to the orientation of ice nanocrystals away from $[110]_c$ zone axis. Sometimes, high density SFs can be observed in all dendrites. c) Random nucleation centers in ice after normalizing the diameter of water droplets. d) The radius-dependent number of nucleation centers follows the same trend as volume, suggesting the nucleation is volume-dependent rather than surface- or thickness-dependent.

## Supplementary Information Fig. S3

**High-speed imaging of the ice crystallization from supercooled water**

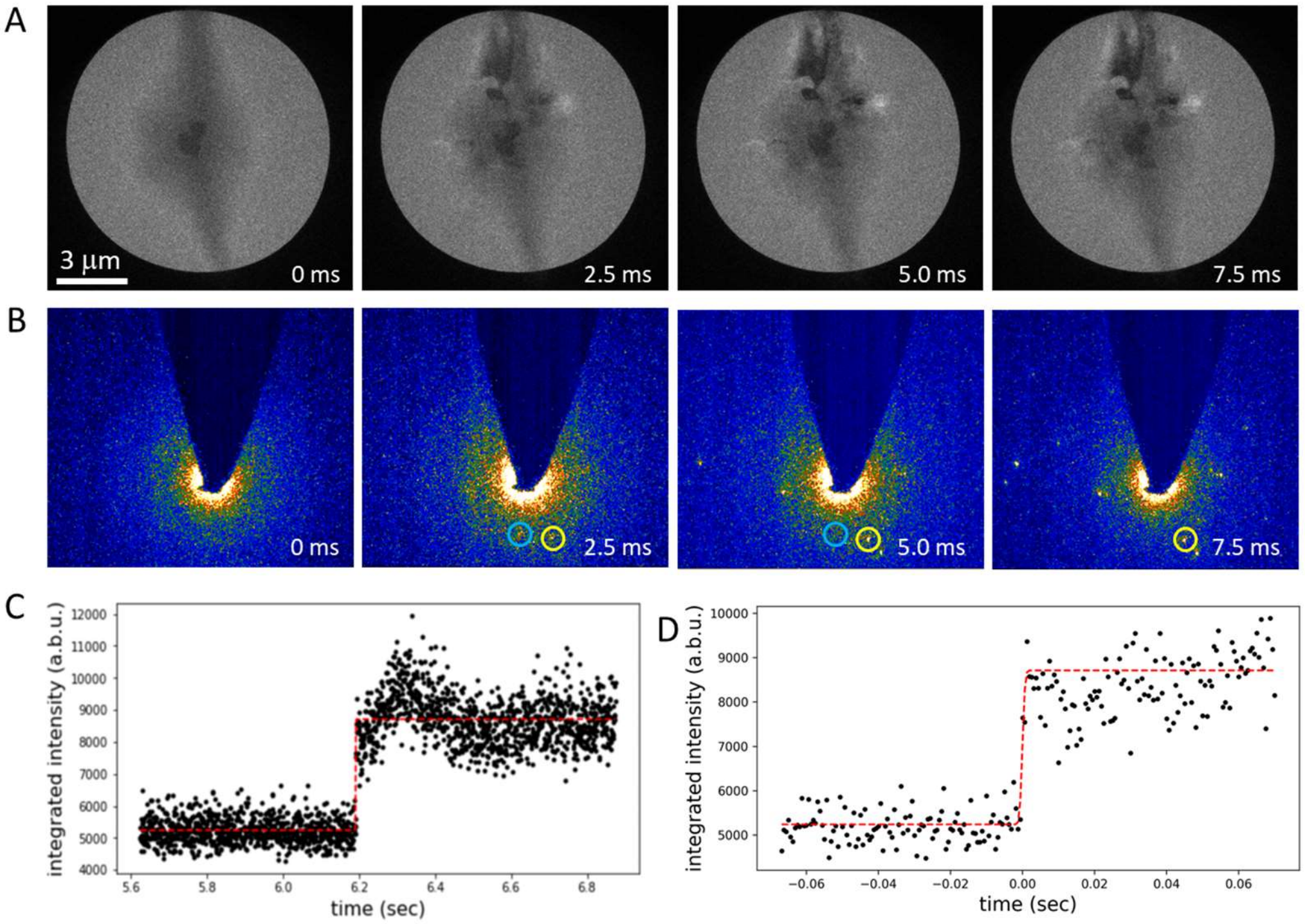

**Fig. S3 High-speed imaging of supercooled water freezing**. a) High speed imaging freezing process of supercooled water using a direct electron detection camera (400 fps). Within one frame, the transformation process is completed, indicating the speed of the crystallization front is higher than 2 mm/s. b) High speed imaging of diffraction pattern of freezing supercooled water using a direct electron detection camera (1600 fps). The intensity of the diffraction spot marked by the yellow circle gradually increases. The diffraction spot marked by the blue circle appeared in a few frames and disappeared after 5 ms. c) The time-dependent integrated intensity of all diffraction spots. d) The time-dependent integrated intensity of all diffraction spots in the 100 ms around the transition. The measured transition time (FWHM) is 3 ms.

## Supplementary Information Fig. S4

**Center-Beam-Blocked Dark-Field (CBBDF) TEM imaging**

We did not observe any significant changes in morphology when using a focused electron beam in the HAADF imaging mode at low magnifications. To avoid any possible beam damage caused by a focused electron beam in the scanning TEM (STEM) mode at high magnification, we applied a board illumination TEM imaging mode, central beam blocked dark-field (CBBDF). The custom-made objective aperture is shown below, blocking the transmission beam. The CBBDF TEM enables imaging ice at a significantly reduced dose rate below 0.5 $e^{-}/nm^{2}s$, while providing Z-contrast similar to HAADF, and allowing the diffraction contrast imaging of defects.

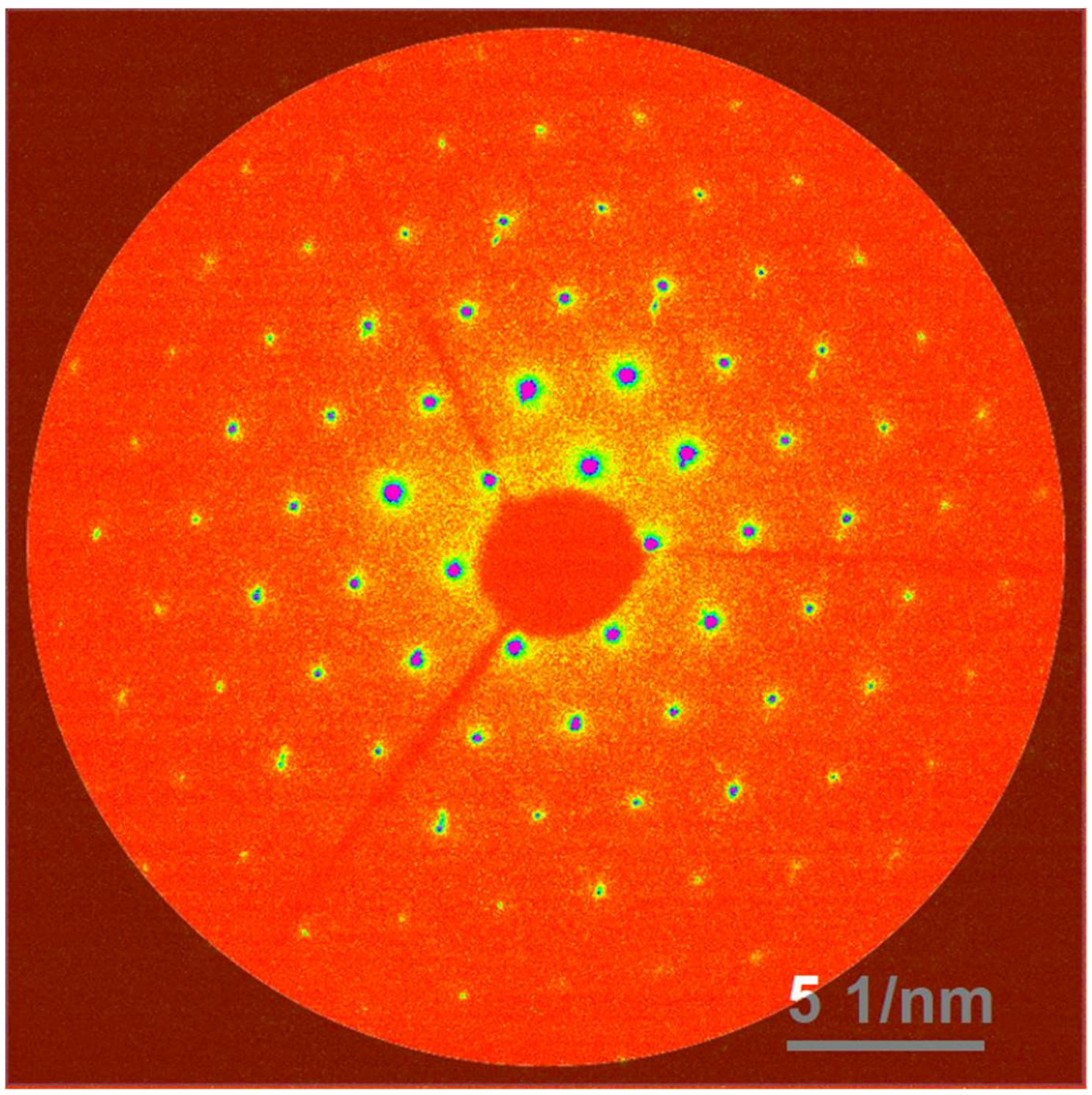

**Fig. S4 A Custom-made objective aperture to realize CBBDF TEM imaging.** A SAED of ice along $[001]_h$ or $[111]_c$ with the CBB objective aperture. CBBDF TEM (Fig. 1g) enables imaging ice at a much lower dose rate, but provides Z-contrast like HAADF, and allows the diffraction contrast imaging of defects. The typical dose rate of the CBBDF imaging is below 50 $e^{-}/Å^{2}s$.

## Supplementary Information Fig. S5

### 3D electron diffraction (3D-ED) of ice frozen from supercooled water

First, a tilt series (-45° to +45°) of 2D nanobeam electron diffraction (NBD) patterns is acquired as schematically shown in Fig. 1H. A 3D diffraction pattern is then reconstructed from the tilt series. The reconstructed 3D electron diffraction pattern enables 360° virtual tilting to reach any orientation beyond the limits of tilt angles of a double-tilt holder. The conventional 2D diffraction pattern (Fig. S5a) is **different** from the projected 3D diffraction pattern (Fig. S5b) in the following: the projected 3D diffraction pattern is obtained by integrating the 3D reciprocal space along the viewing direction, unlike the 2D diffraction pattern which is a section of the 3D reciprocal space. The 3D-ED pattern (Fig. S5e) of all dendrites including a tertiary dendrite matches with the simulated 3D-ED pattern of Ice $I_{sd}$, indicating tertiary dendrites consist of Ice $I_{sd}$.

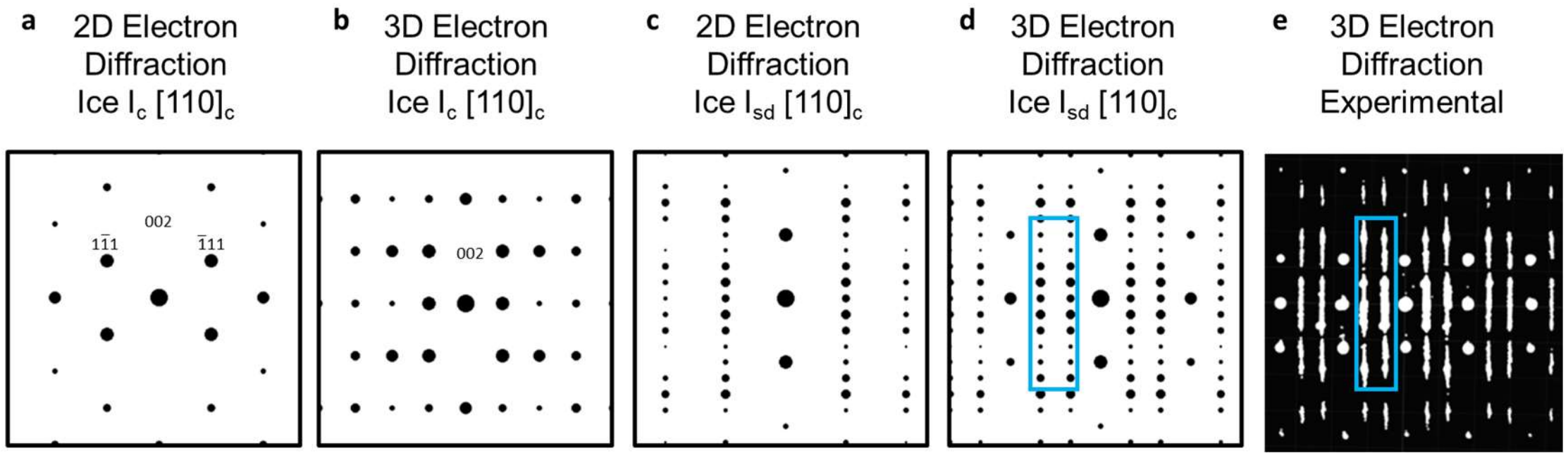

**Fig. S5 Projected 3D electron diffraction patterns.** a) A simulated conventional diffraction pattern of $I_c$ along $[110]_c$. b) A simulated projected 3D diffraction pattern of $I_c$ along $[110]_c$. c) A simulated conventional diffraction pattern of $I_{sd}$ along $[110]_c$. d) A simulated projected 3D diffraction pattern of $I_{sd}$ along $[110]_c$. Note two extra rows of diffractions spots appear as indicated by the box. e) An experimental projected 3D diffraction pattern of tertiary $I_{sd}$ along $[110]_c$, matching with the projected 3D pattern in d).

## Supplementary Information Fig. S6

### Cubic order of the ice frozen from supercooled water

Three-dimensional electron diffraction (3D-ED) patterns from individual dendritic grains exhibit pronounced streaking arising from a high density of stacking faults, confirming their $I_{sd}$ nature. In contrast, 3D-ED patterns acquired from pairs of primary and tertiary dendrites reveal a fixed crystallographic misorientation of 109.5°, demonstrating that all branches are related by twinning on {111}c planes. When diffraction data from all branches within a single ice crystal are combined, the reconstructed 3D-ED volume resembles that of a simple cubic lattice, displaying fourfold symmetry along the [100] direction (Fig. 1L) and threefold symmetry along the [111] direction (Fig. S6). Collectively, the eight dendrites occupy the eight octahedral facets of a cubic lattice (Fig. 1I), resulting in a globally cubic-symmetric morphology despite the pronounced stacking disorder within each individual branch.

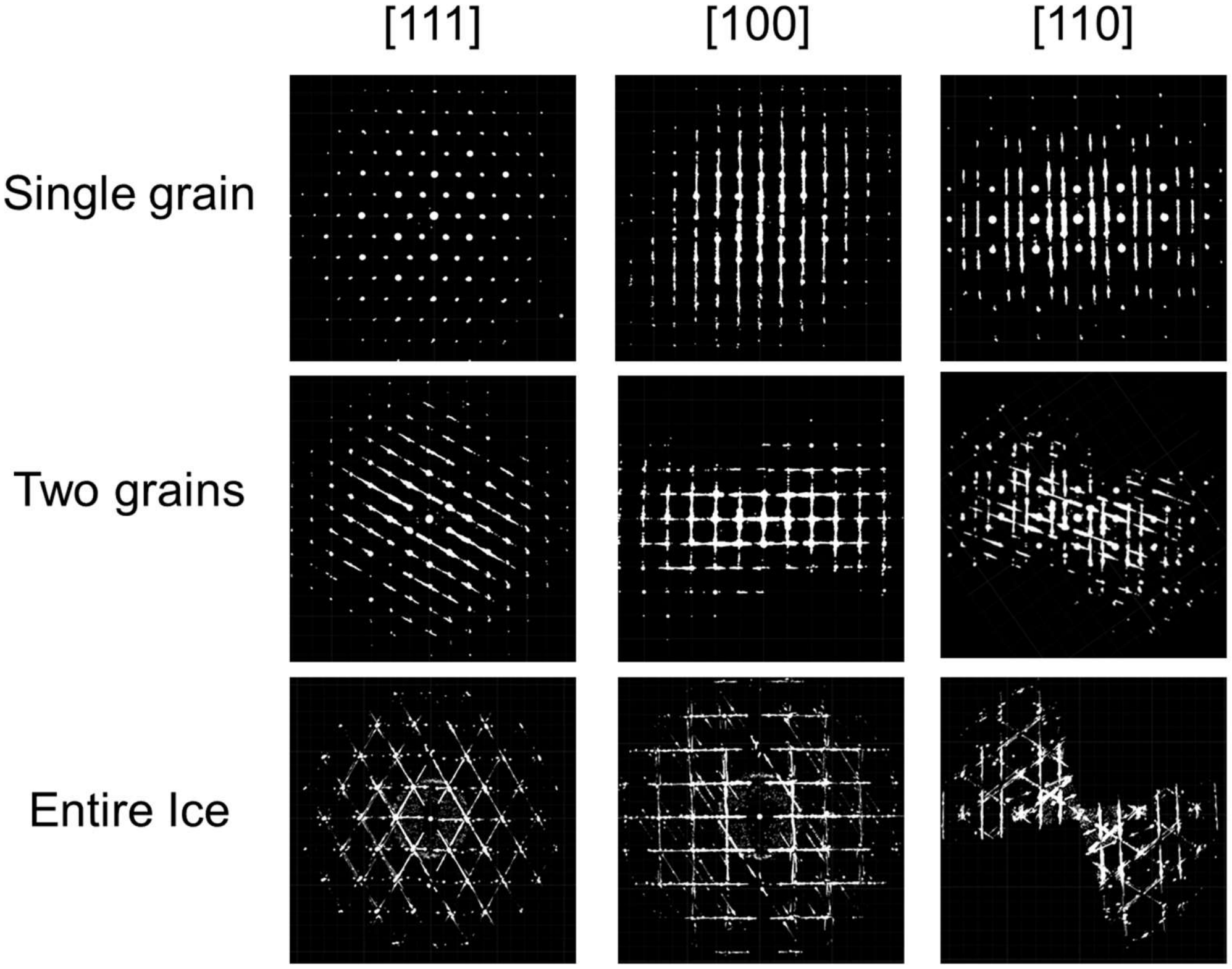

**Fig. S6 Projected 3D electron diffraction patterns.** The 3D-ED patterns acquired from the entire ice along the [111], [001], and [110] zone axes exhibit diffraction features characteristic of cubic symmetry.

## Supplementary Information Fig. S7

**Cubicity measurements at the nucleation center and dendrites**

As shown in Figs. S6a-c, both ice $I_h$ and $I_c$ contain identical layers of hydrogen-bonded water molecules, only differing in their stacking sequences[2]. Atomic models of stacking sequences of ABAB in ice $I_h$ along $[100]_h$ (Fig. S7a) and stacking sequences of ABC in ice $I_c$ along $[110]_c$ (Fig. S7b). Local horizontal mirror planes marked by the green line (with H on the right side) can be found ice $I_h$ and local inversion centers marked by blacked dots (with K) can be found in ice $I_c$. In ice $I_{sd}$ (Fig. S7c), hexagonal and cubic ice segments are defined by the local symmetry H and K[3,4]. Using the local symmetry, we can precisely determine the stacking sequence from a HRTEM image. Figs. S7d-f show a low-dose HRTEM image from a large portion of the nucleation core, secondary, and a tertiary from another area acquired with a direct detection camera (Small portion are shown in Fig. 2). By applying the image processing to locate atomic columns in the HRTEM image, computer decoded ice $I_h$ (red color) and ice $I_c$ (blue color) segments can be analyzed from the measured stacking sequence based on the definition above.

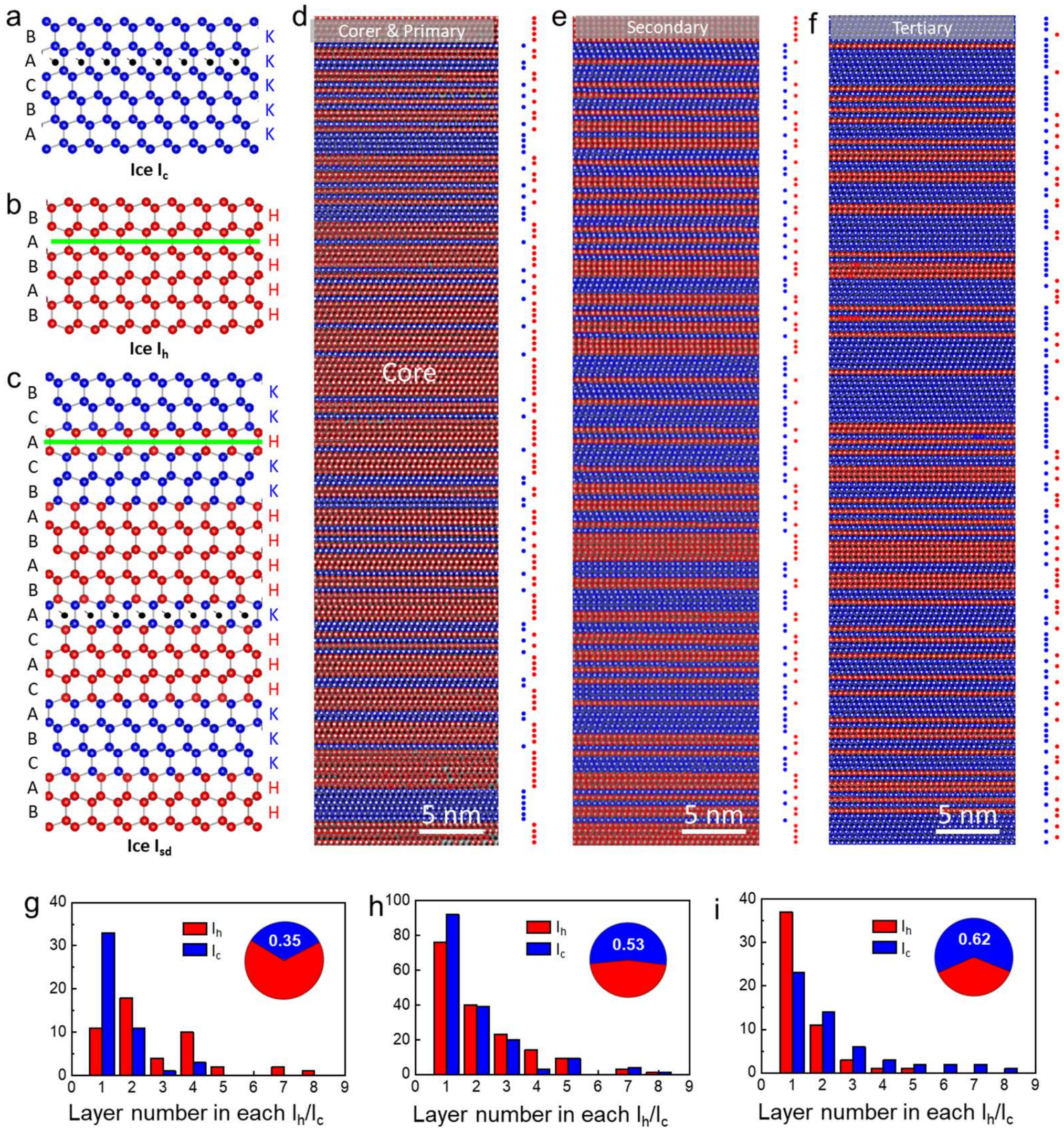

**Fig. S7 Local cubicity measurements from low-dose HRTEM images of ice.** Atomic models showing stacking sequences of ABAB in ice $I_h$ along [100] (a) and stacking sequences of ABC in ice $I_c$ along $[110]_c$ (b). Local horizontal mirror planes marked by the green line (with H on the right side) can be found ice $I_h$ and local inversion centers marked by blacked dots (with K) can be found in ice $I_c$. In ice $I_{sd}$ (c), hexagonal and cubic ice segments are defined by the local symmetry H and K. The stacking sequence can be determined using the relative atomic column locations in the HRTEM image. Ice $I_h$ (red color) and ice $I_c$ (blue color) segments can be analyzed from the measured stacking sequence based on the definition. d) Low-dose HRTEM image from the nucleation core and the computer-decoded ice $I_h$ segments (red dots) and ice $I_c$ segments (blue dots). Cubicity can be directly measured using this method as 0.35 from the area between tips of two secondary dendrites. e) A section of low-dose HRTEM image from a secondary dendrite and the corresponding computer-decoded ice $I_h$ and ice $I_c$ segments, showing

the cubicity of 0.53. f) A section of low-dose HRTEM image from a tertiary dendrite and the corresponding computer-decoded ice $I_h$ and ice $I_c$ segments, showing the cubicity of 0.62. g) The ice $I_{sd}$ at the core has a cubicity of 0.35. Many single layers of ice $I_c$ (bule) appear as SFs in the ice $I_h$ (red) segments as shown from the layer number distribution of each ice $I_c$ or $I_h$ segments. Primary dendrites have the same cubicity as the core. h) The ice $I_{sd}$ in the secondary dendrites have a cubicity of 0.53. The probability of monolayer ice $I_h$ (red) and ice $I_c$ (blue) in the regions is almost equal. i) The ice $I_{sd}$ in the tertiary dendrites have a cubicity of 0.62. Many single layers of ice $I_h$ (red) appear as SFs in the ice $I_c$ (blue) segments as shown from the layer number distribution of each ice $I_c$ or $I_h$ segments.

## Supplementary Information Fig. S8

**Atomic structures, HRTEM images and diffraction patterns for Ice $I_h$, $I_c$, and $I_{sd}$.**

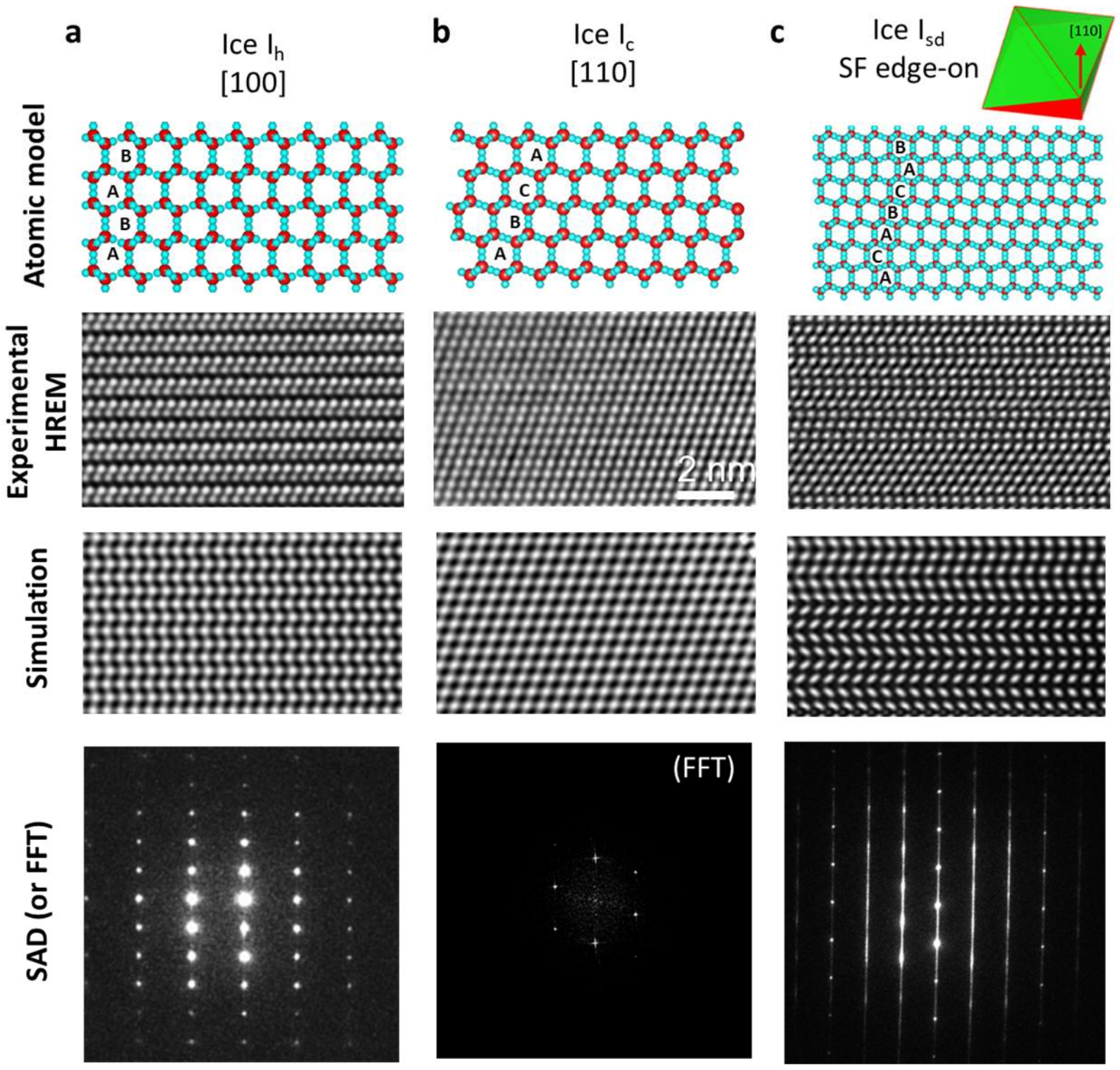

**Fig. S8 Atomic structures, HRTEM images and diffraction patterns for Ice $I_h$, $I_c$, and $I_{sd}$.** a) Ice $I_h$. Stacking sequence ABAB can be observed along $[100]_h$. b) Ice $I_c$. Stacking sequence ABC can be observed along $[110]_c$ at local areas. Since no large grain of pure Ice $I_c$ is observed, a fast Fourier transform (FFT) pattern from a small area of ice $I_c$ is used as diffraction pattern. c) ice $I_{sd}$ when stacking planes are edge-on. In this case (for example in primary and secondary dendrites), stacking sequence can be clearly observed along $[100]_h$ or $[110]_c$. A vertical streaking pattern perpendicular to the horizontal stacking planes can be observed in the SAD pattern.

## Supplementary Information Fig. S9

**Size distribution of cubic germs**

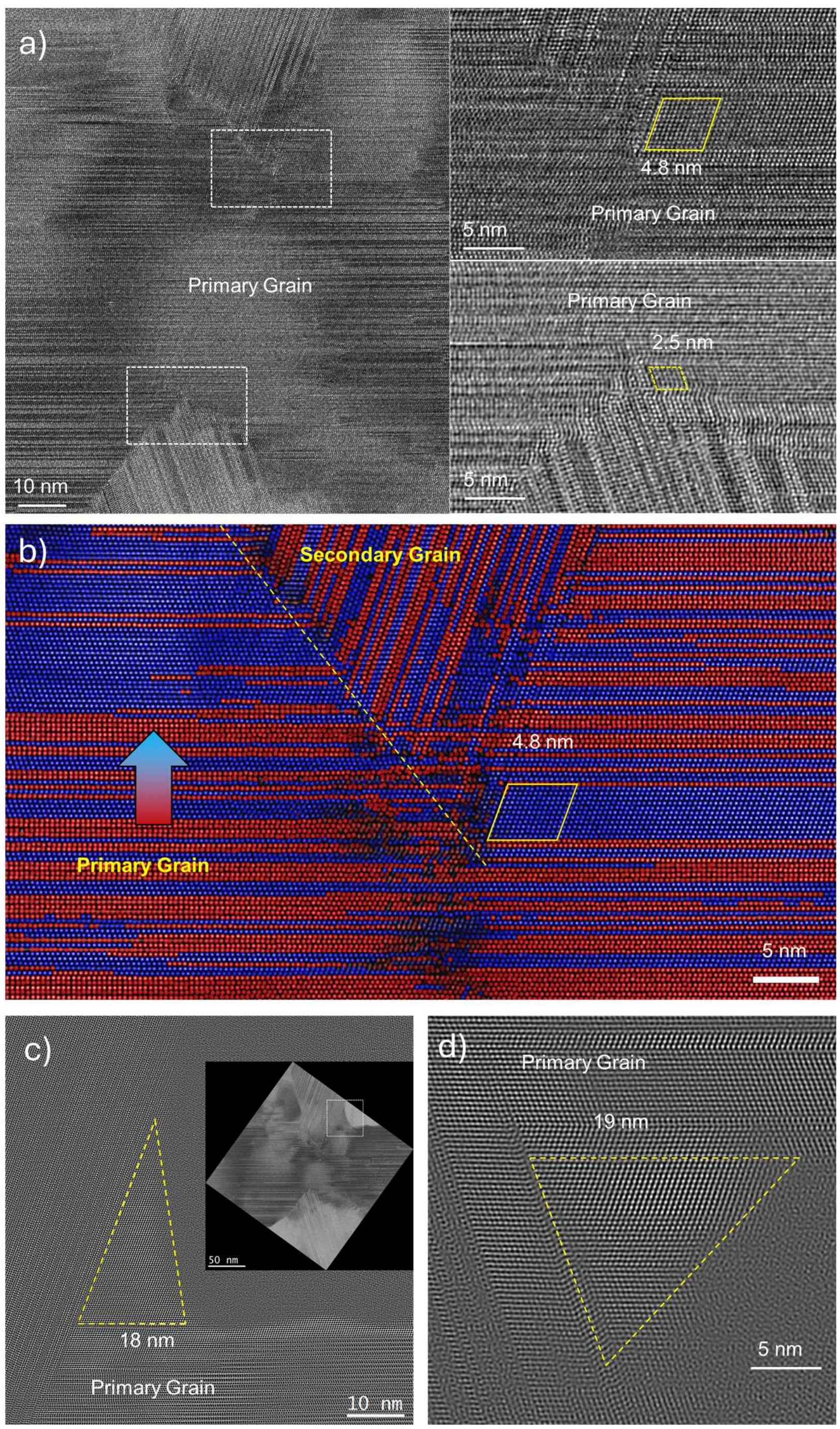

**Fig. S9 Size distribution of cubic ice germs.** a) and b) cubic ice germs at the tip of secondary dendrites. c) and d) cubic ice germs at the tip of tertiary dendrites. The size of cubic ice germs is smaller ranging from 2.5 nm to 13 nm, while the size of cubic ice germ is around 20 nm

## Supplementary Information Fig. S10

**Molecular dynamics of the nucleation of cubic ice germ**

**MD simulations details:**

We initiated the simulations with ABCACB stacked Ice configurations to create the initial setup as depicted in Fig. S10. The liquid configurations were generated by supercooling liquid water. We utilized the atomsk software[5] in conjunction with the Pymatgen python library[7] to create the combined setup with crystals of stacking disordered ice ($I_{sd}$) and hexagonal ice ($I_h$) at the top and bottom, with a liquid region in between. To obtain the top $I_{sd}$ configuration, we rotated the c axis of the ABCACB ice by $70.5^0$ degrees with respect to the $[110]_c$ facet of the bottom $I_{sd}$. The LAMMPS package[6] with a coarse-grained bond order potential[7] was employed for the simulations. The equations of motion were integrated using the velocity-Verlet algorithm. The dimensions of the simulation box were approximately 120 Å x 120 Å x 110 Å, with the width of the crystalline sections being around 20 Å along the z direction. The initial configurations underwent equilibration at 200K using an NVE ensemble with a timestep of 2 fs for approximately 0.4 ns. During the equilibration, the crystalline sections were kept intact. The equilibrated configurations were then simulated at temperatures of 230K, 240K, 250K, and 260K, respectively, using an NVE ensemble. The temperatures were maintained using a Langevin thermostat, and external stress was applied along the normal to crystalline section to maintain a pressure of 1 bar. For the temperature simulations, a time step of 10 fs was used. To maintain the twinning relationship while growing through the liquid, the top and bottom crystalline parts with a width of 18 Å were kept fixed throughout the simulations. For the characterization of particles belonging to the hexagonal and cubic motif in simulation trajectories, a modified common neighbor analysis (CNA)[8] was used.

**Formation of twin boundary of growth along the prism planes of stacking disordered ICE ($I_{sd}$):**

We also investigated the formation of twin boundaries, where both the top and bottom sections had their prism planes exposed to the liquid (Fig. S10(e-g)). Since the growth at the prism plane facet is more prominent for both the top and bottom crystalline sections, we did not observe any nucleation of cubic germ in this case. This observation aligns with the fact that there is no cubic twinning relationship in this scenario. The ratio of hexagonal ice to cubic ice during the simulations at different temperatures remained approximately 0.5, as shown in Fig. S10(g). This indicates a uniform growth along the facets throughout the simulation.

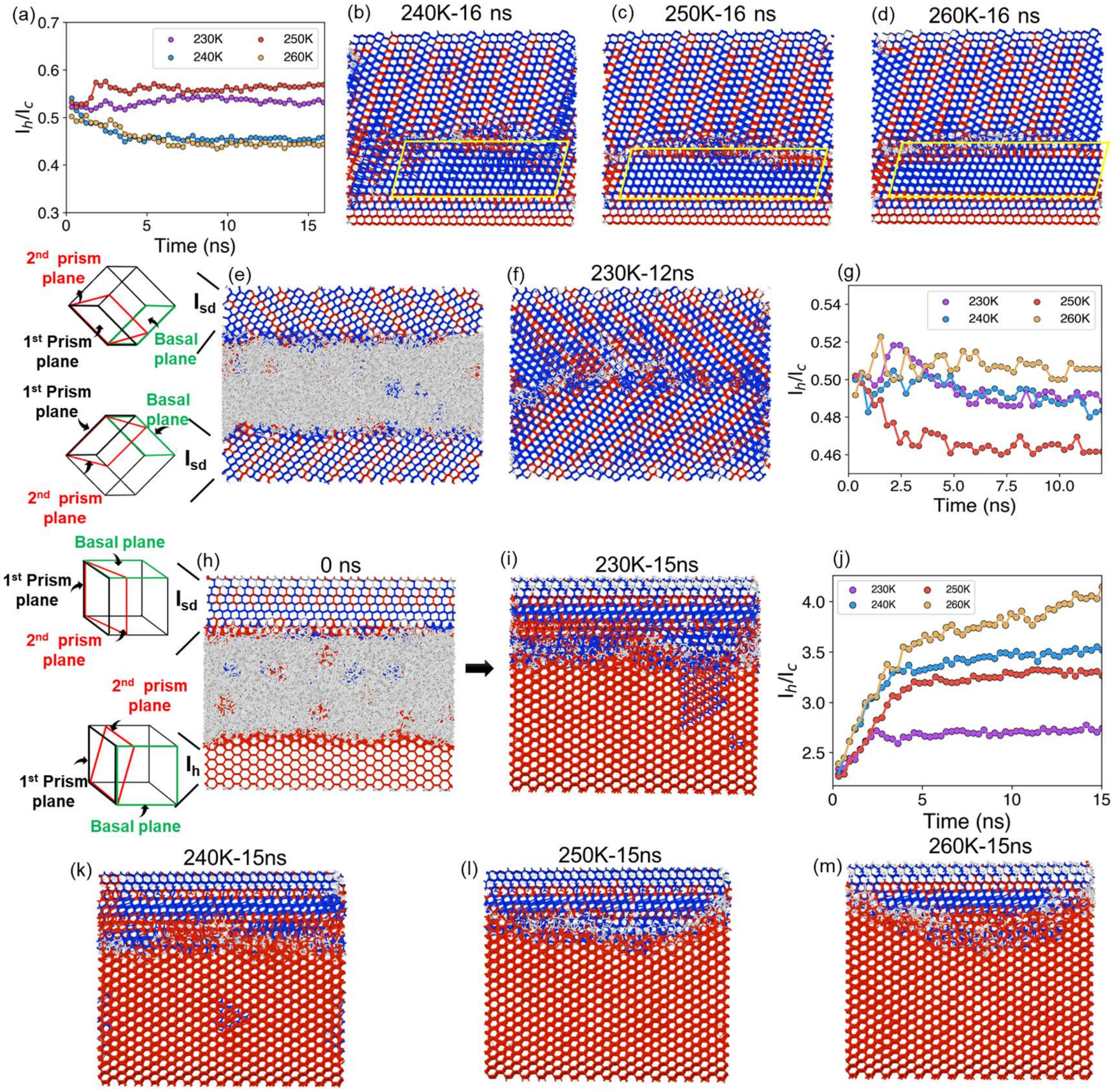

Fig. S10(a) illustrates the evolution of the ratio of hexagonal and cubic ice over time during the nucleation of a cubic germ and the formation of twin boundaries in stacking disordered ice ($I_{sd}$).

Figures (b-d) present the final frames capturing the nucleation and growth of the cubic germ (highlighted in yellow) at temperatures 230K, 250K, and 260K, respectively, at times 8ns, 12ns, and 12ns. (e) shows the initial setup for the growth of the twin boundary in $I_{sd}$, where both the top and bottom crystals have their prism planes exposed to the liquid. In Figure (f), a frame from the simulation after 12ns at 230K is displayed. (g) demonstrates the evolution of the ratio of hexagonal and cubic ice in the system over time at different temperatures, corresponding to the setup depicted in (e). Finally, Figures (h-m) depict the final frames of the simulations at different temperatures, clearly illustrating the increasing dominance of ice $I_h$ with rising temperature.

**MD simulation results of the dominance of ice $I_h$ near $T_m$:**

In the context of ice formation, the appearance of ice $I_h$ at the edges is a notable observation in experiments. It is consistent with Ostwald's rule, which suggests that a metastable ice phase (ice $I_{sd}$) nucleates first. This is possible because the basal plane of hexagonal ice and the {111} plane of cubic ice is equivalent, allowing for a seamless stacking of cubic and hexagonal layers[5]. Over time, the presence of cubicity gradually decreases with annealing, eventually approaching pure ice $I_h$[1, 5]. To observe this effect near twin boundary dendritic morphologies, we performed MD simulations in a similar setup as shown in Fig. 3E, however, in our simulations, the bottom layer is hexagonal ice. Additionally, it has been observed that hexagonal ice growth exhibits anisotropic behavior [3, 4, 5].Specifically, the growth and melting rates of the basal face are slower compared to the other two faces, in comparison to the prism faces[2]. To observe the evolution of cubic and hexagonal ice within the timescales of the MD simulations, we created a stacking disordered ($I_{sd}$) top section, with its basal plane exposed to the liquid. From Fig. S10(j), it is evident that the fraction of hexagonal ice ($I_h$) increases considerably with increasing temperature, inhibiting the growth of stacking disordered ice (or the amount of cubic ice in the system). This observation is further supported by the final snapshots of the configuration after the MD simulations. At 230K, the growth of $I_{sd}$ and some remaining cubic ice (blue color) can be seen within the hexagonal ice (red color). However, there is a consistent decrease in $I_{sd}$ content for temperatures 240K, 250K, and 260K (Fig. S10(h-m)).

## Supplementary Information Fig. S11

## Numerical simulation of freezing of supercooled micro-sized water droplet

### SI-11.1 Physical model for freezing of supercooled micro-sized water droplet

In the experiments of freezing of supercooled micro-sized water droplets, the water droplets were separately encapsulated by two carbon films into flat ellipsoidal shape and the dimensions of single water droplet are about 0.5-5 μm in radius and 30-250 nm in maximum thickness, as shown in Fig. S11a. In the numerical simulation, the geometry of the flat ellipsoidal-shaped water droplet is reasonably simplified into the disc-shaped since the diameter-height ratio of water droplet is extremely large, as shown in Fig. S11b. In addition, the water droplet cannot fully adhere to carbon film due to its hydrophobicity and the contact angle of water droplet on the carbon film is large. In the simulation, only 4 % of the outer surface of water droplet is considered to contact with the carbon film and the contact surfaces are on the top and bottom sides of water droplet since the water droplet is held by two carbon films, as shown in Fig. S11b. In addition, the carbon film is connected to the cryogenic transmission electron microscopy where the temperature is maintained at -39℃ at first and suddenly dropped to -40℃. Therefore, the initial temperature of water droplet is -39℃ and boundary conditions of the contact surfaces are maintained at the constant temperature of -40℃. The non-contact surfaces are considered to be adiabatic.

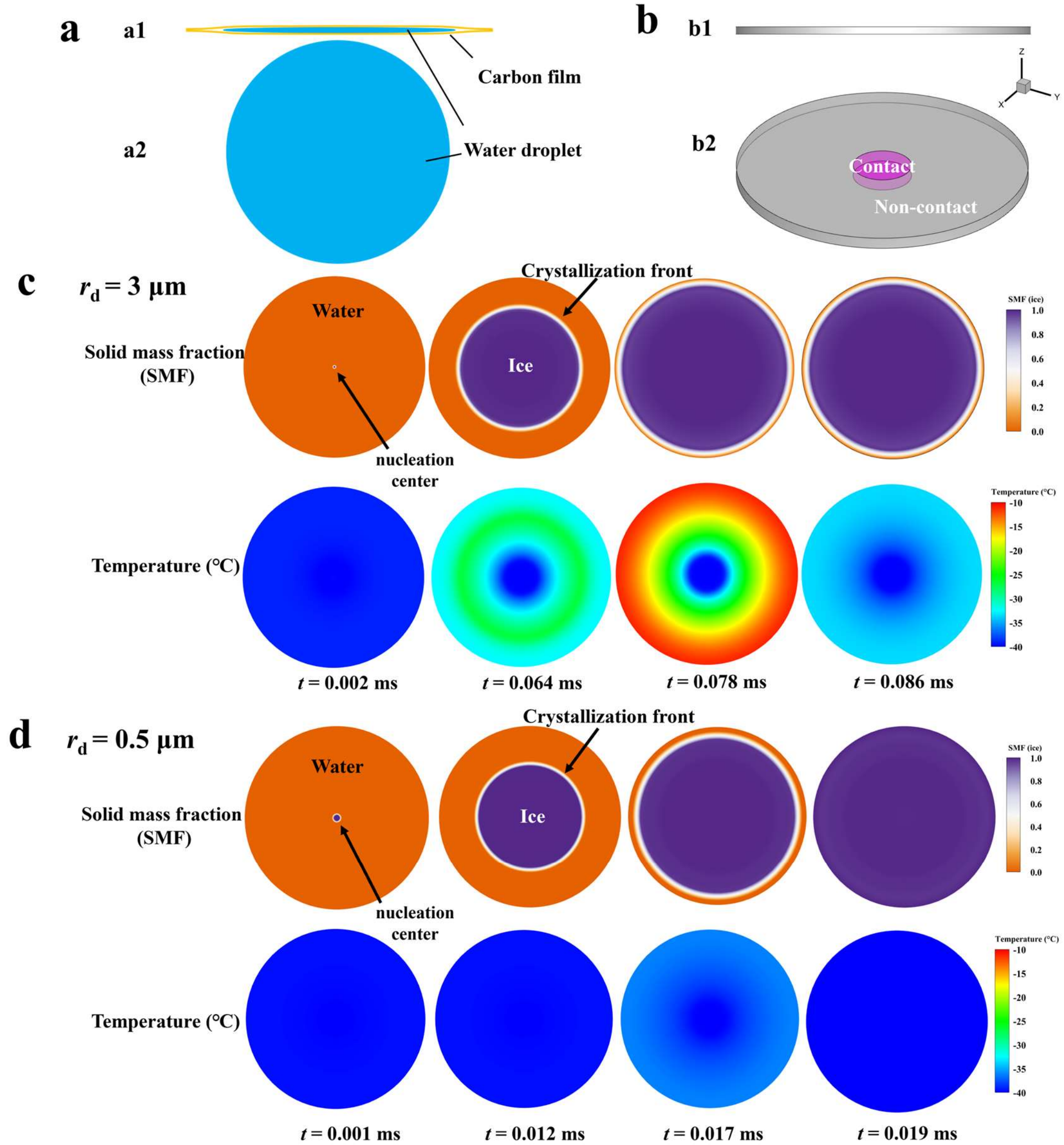

**Fig. S11 Numerical simulation of freezing of supercooled micro-sized water droplet**. a) The flat ellipsoidal-shaped micro-sized water droplet a1) side view; a2) bird view. b) The disc-shaped geometry of water droplet for numerical simulation b1) side view; b2) general view. The evolutions of solid mass fraction (ice) and temperature distributions during the freezing of water droplet with the radius of c) 3 μm and d) 0.5 μm.

### SI-11.2 Numerical model for freezing of supercooled micro-sized water droplet

Based on the physical model, the numerical model is proposed to simulate the freezing of supercooled micro-sized water droplet. During the entire freezing of supercooled micro-sized water droplet, there are four stages including the water-cooling stage, recalescence stage, solidification stage and ice-supercooling stage. The cooling stage of water, solidification stage and ice-supercooling stage can be accurately described by thermodynamic model. However, the recalescence stage cannot be simply described by thermodynamic model since crystallization kinetics come into play in this highly transient process. Therefore, we established a 3-dimensional numerical model which considers both the crystallization kinetics and thermodynamics to accurately describe the freezing dynamics of supercooled micro-sized water droplet[9,10].

#### SI-11.2.1 Thermodynamic model

In the numerical model, the natural convection caused by the temperature gradient of water is neglected since the intensity of natural convection is negligible for micro-sized water droplet. Therefore, the energy equation of water droplet can be depicted by the enthalpy method.

$$\frac{\partial}{\partial t}[\chi \rho_{\mathrm{d}} H_{\mathrm{i}} + (1-\chi)\rho_{\mathrm{d}} H_{\mathrm{w}}] = \nabla \cdot (k_{\mathrm{d}} \nabla T_{\mathrm{d}}) \quad \text{(S1)}$$

Where, $\chi$ refers to the mass fraction of ice and $t$ is the time. $\rho_{\mathrm{d}}$, $k_{\mathrm{d}}$ and $T_{\mathrm{d}}$ refer to the density, thermal conductivity and temperature of water droplet, respectively. $H_{\mathrm{i}}$ and $H_{\mathrm{w}}$ are the enthalpy of ice and water which can be formulated as follows.

$$H_{\mathrm{i}} = H_{\mathrm{ref,i}} + \int_{T_{\mathrm{ref}}}^{T} C_{\mathrm{pi}} dT \quad \text{(S2)}$$

$$H_{\mathrm{w}} = H_{\mathrm{ref,w}} + \int_{T_{\mathrm{ref}}}^{T} C_{\mathrm{pw}} dT \quad \text{(S3)}$$

The reference enthalpies of ice and water, i.e., $H_{\mathrm{ref,i}}$ and $H_{\mathrm{ref,w}}$ follow the relationship $H_{\mathrm{ref,w}} = H_{\mathrm{ref,i}} + L$ and $L$ refers to the latent heat of water. $C_{\mathrm{pi}}$ and $C_{\mathrm{pw}}$ are the thermal capacities of ice and water, respectively.

#### SI-11.2.2 Crystallization kinetic model

In the recalescence stage, the ice nucleation is first formed in supercooled water droplet and then ice crystal grows outward from the nucleation center. For the micro-sized water droplet, the nucleation is considered homogeneous and the nucleation rate $J$ can be formulated as follows[10].

$$J = \frac{A_{\mathrm{m}}}{3}\left(\frac{\rho_{\mathrm{w}} N_{\mathrm{A}}}{M}\right)^{7/3} \sqrt{\frac{A_{\mathrm{m}}\sigma}{\pi k_{\mathrm{B}} T}} D \times \exp\left[-\frac{4\left(\frac{A_{\mathrm{m}}\sigma}{k_{\mathrm{B}}T}\right)^{3}}{27\left(\ln \frac{P_{\mathrm{sw}}}{P_{\mathrm{si}}}\right)^{2}}\right] \quad \text{(S4)}$$

The detailed information of the parameters in equation (S4) can be found the previous study[9]. Based on the nucleation rate $J$, the quantity of the ice nucleus $N_{\mathrm{v}}$ in water droplet can be obtained.

$$N_{\mathrm{v}}(t) = \int_{0}^{t} [V_{\mathrm{d}} - V_{\mathrm{i}}] J(t) \mathrm{d}t \quad \text{(S5)}$$

Where, $V_{\mathrm{i}}$ and $V_{\mathrm{d}}$ refer to the volumes of ice and water droplet, respectively. $N_{\mathrm{v}}$ determines whether the water droplet starts to freeze. When the $N_{\mathrm{v}} < 1.0$, ice nucleus is forming; Once $N_{\mathrm{v}} = 1.0$, the ice nucleus is formed and immediately the ice crystal grows outward from the nucleation center. The propagating velocity $v_{\mathrm{g}}$ of ice crystal is temperature-dependent[11,12].

In order to accurately describe the growth of ice crystal, the Level-Set method is introduced to capture the interface of the crystallization front of water droplet, and the Level-Set function Φ is defined as the distance function.

$$\frac{\partial \Phi}{\partial t} + v_{\mathrm{g}} \Delta \Phi = 0 \quad \text{(S6)}$$

The Φ of grid is positive in water zone and is negative in ice zone. The interface where Φ=0 indicates the crystallization front of water droplet.

**SI-11.3 Temporal evolutions of the temperature and solid mass fraction (SMF) distributions**

In supercooled micro-sized water droplet, the temperature and solid mass fraction distributions are nearly uniform in the Z-direction due to small thickness of water droplet, as shown in Fig. S11a and S11b. We take the horizontal cross section (XOY plane) at the center of water droplet for illustration. The time instant that the supercooled water droplet starts to freeze is taken as $t$=0, and the distributions of temperature and solid mass fraction (SMF) on XOY plane of supercooled water droplet with the radius $r_d$ of 3 μm and 0.5 μm at different time instants during freezing are depicted in Fig. S11c and S11d, respectively. It is found that the solid mass fraction near the crystallization front increases from 0 to 1 within a narrow transition zone, indicating the release of latent heat of fusion and therefore, the temperature near the crystallization front is the highest. When the crystallization front approaches the edge of water droplet, i.e., about $t$ = 0.078 ms for $r_d$ = 3 μm and about $t$ = 0.017 ms for $r_d$ = 0.5 μm, the temperature gradient in the water droplet is the most significant during freezing. At $t$ = 0.078 ms for $r_d$ = 3 μm, it is found that the temperature is the highest near the edge of water droplet, indicating a largest temperature increase since it is farthest from cooling source and the temperature is the lowest near the center of water droplet, indicating the smallest temperature increase due to the shortest distance from the cooling source.